\documentclass[a4paper,12pt]{article}
\usepackage{jheppub}
\usepackage{times}
\usepackage{xcolor}
\usepackage{graphicx} 
\usepackage{microtype}
\usepackage{float} 
\usepackage{caption}
\usepackage{booktabs}
\usepackage{algorithm}
\usepackage{algorithmic}

\def\beq{\begin{equation}}
\def\eeq{\end{equation}}
\def\bea{\begin{eqnarray}}
\def\eea{\end{eqnarray}}
\def\beqn{\begin{eqnarray}} 
\def\eeqn{\end{eqnarray}}

\def\Eq#1{Eq.~(\ref{#1})}

\def\qon#1{q_{#1,0}^{(+)}}

\def\ket#1{|{#1}\rangle}

\def\qb{\mathbf{q}}

\pdfoutput=1

\newcommand{\valencia}{Instituto de F\'{\i}sica Corpuscular, Universitat de Val\`{e}ncia 
-- Consejo Superior de Investigaciones Cient\'{\i}ficas, 
Parc Cient\'{\i}fic, E-46980 Paterna, Valencia, Spain.}
\newcommand{\culiacanA}{Facultad de Ciencias F\'{\i}sico-Matem\'aticas,
Universidad Aut\'onoma de Sinaloa, Ciudad Universitaria, CP 80000 Culiac\'an, Mexico.}

\begin{document}

\title{Graph theory-based automated quantum algorithm for efficient querying of acyclic and multiloop causal configurations}

\author[a,b]{Salvador~A.~Ochoa-Oregon,}
\author[b]{Juan~P.~Uribe-Ramírez,}
\author[b]{Roger~J.~Hernández-Pinto,}
\author[b]{Selomit~Ramírez-Uribe,}
\author[a]{and~Germ\'an~Rodrigo}

\affiliation[a]{\valencia}
\affiliation[b]{\culiacanA}

\emailAdd{salvadorochoa.fcfm@ms.uas.edu.mx}
\emailAdd{juanuribe.fcfm@uas.edu.mx}
\emailAdd{roger@uas.edu.mx}
\emailAdd{selomitru@uas.edu.mx}
\emailAdd{german.rodrigo@csic.es}

\abstract{Quantum algorithms provide a promising framework in high-energy physics, in particular, for unraveling the causal configurations of multiloop Feynman diagrams by identifying Feynman propagators with qubits, a challenging task closely analogous to querying \textit{directed acyclic graphs} in graph theory. 
In this paper, we present the Minimum Clique-optimised quantum Algorithm (MCA), an automated quantum algorithm designed to efficiently query the causal structures within the Loop-Tree Duality. The MCA quantum algorithm is optimised by exploiting graph theory techniques, specifically, by analogy with the Minimum Clique Partition problem. The  performance of the MCA quantum algorithm across diverse multiloop topologies is assessed by analysing the transpiled quantum circuit depth and quantum circuit area, showing a significant reduction in the quantum resources needed for implementation.}

February 19, 2026

\maketitle

\section{Introduction and motivation}
\label{sec:Introduction}
The description of the fundamental interactions of elementary particles in High-Energy Physics (HEP) requires a deep understanding of Quantum Field Theory (QFT). Large-scale collider experiments, most notably the CERN's Large Hadron Collider (LHC), are devoted to rigurously testing the predictions of the Standard Model (SM) with unprecedented accuracy, in the ongoing quest for signals of new physics beyond the current theoretical framework. 

The central bottleneck of theoretical HEP is the treatment of quantum fluctuations at increasingly high orders in the perturbative expansion. Achieving precise theoretical predictions hinges on the delicate cancellation of ultraviolet (UV) and infrared (IR) singularities that are intrinsic to scattering amplitudes in QFT. These singularities proliferate dramatically as the number of loops and external particles increases. In perturbative Quantum Chromodynamics (pQCD), the procedure for obtaining finite theoretical predictions at one loop is well established. Beyond first order, however, the cancellation of divergences becomes increasingly tedious and technically intricate.

A landmark success of perturbative methods was the phenomenological description of the Higgs boson, which resulted in its discovery in 2012~\cite{ATLAS:2012yve,CMS:2012qbp}. The importance of this fundamental particle in the SM motivated calculations at Next-to-Next-to-Leading Order (NNLO)~\cite{Catani:2001ic,Harlander:2002wh,Anastasiou:2002yz} in the strong coupling. Recent advances have enhanced precision, reaching N$^3$LO for key LHC processes such as gluon-fusion Higgs production~\cite{Catani:2014uta,Anastasiou:2014vaa,Mistlberger:2018etf,Billis:2021ecs,Chen:2021isd} and Higgs pair production with N$^3$LO+N$^3$LL resummation~\cite{AH:2022elh}, yielding sub-percent scale uncertainties. These developments include NNLO predictions for many other processes, enabling detailed jet substructure and flavor analyses, alongside fully differential NNLO distributions for high-$p_T$ probes of new physics. Such progress is vital for matching the expected experimental precision of the High-Luminosity LHC, improving electroweak precision observables, strong coupling determinations, and beyond-SM sensitivity, while laying the groundwork for future colliders like the Future Circular Collider (FCC)~\cite{FCC:2025lpp} or the Linear Collider~\cite{LinearColliderVision:2025hlt} through innovative handling of multiscale Feynman integrals, threshold resummation and numerical methods.

Methods for computing multiloop scattering amplitudes in perturbative QFT, especially pQCD, centre on reducing complex Feynman integrals to a small set of master integrals (MIs) via integration-by-parts (IBP) identities~\cite{Grozin:2011mt}, followed by evaluation using analytic or numerical techniques. Key approaches include the Laporta algorithm for linear IBP reduction~\cite{Laporta:2000dsw}, differential equations for solving MIs~\cite{Henn:2013pwa}, numerical unitarity for extracting coefficients~\cite{Abreu:2020xvt}, sector decomposition~\cite{Binoth:2003ak,Smirnov:2008py} and Mellin-Barnes representations~\cite{Gluza:2007rt}. Recent innovations, such as generating functions yielding symbolic recurrence relations~\cite{Smirnov:2003kc}, intersection theory~\cite{Mastrolia:2018uzb}, finite-field~\cite{Klappert:2020nbg}, and  Gr\"obner bases~\cite{Smirnov:2006tz}, address the exponential growth in complexity for high-rank, multiloop topologies, enabling results like two-loop five-parton QCD amplitudes or three-loop processes. Despite the progress developed so far, achieving the theoretical precision required at high-energy colliders calls for exploring approaches that surpass the current state of the art and drive further progress towards higher perturbative orders. 

The Loop-Tree Duality~(LTD)~\cite{Catani:2008xa,Bierenbaum:2010cy,Bierenbaum:2012th,Runkel:2019yrs,Capatti:2019ypt,Verdugo:2020kzh,Soper:1999xk} has proven its capabilities to localize  IR divergences within a finite region of the integration domain~\cite{Buchta:2014dfa,Buchta:2015wna,Aguilera-Verdugo:2019kbz}, while clearly interpreting the emergence of UV and IR singularirites in terms of causality~\cite{snowmass2020,Aguilera-Verdugo:2020kzc,Ramirez-Uribe:2020hes,Tomboulis:2017rvd,Sborlini:2021owe,Imaz:2025buf}. The relevance of the LTD framework is founded on the intrinsic property of providing more stable integrands by manifestly cancelling all non-causal singularities. Furthermore, it has been recently shown that vacuum amplitudes in LTD~\cite{Ramirez-Uribe:2024rjg, LTD:2024yrb}, i.e. scattering amplitudes without external particles, provide integrand representations of physical observables that are well-defined directly in the four physical dimensions of the spacetime. This approach significantly reduces the complexity with respect to other four-dimensional implementations~\cite{Hernandez-Pinto:2015ysa,Sborlini:2016hat,Sborlini:2016gbr,Prisco:2020kyb,Capatti:2020xjc}, which in a first step requires casting all causal configurations of multiloop vacuum diagrams, thereby offering a physics-motivated roadmap for QFT calculations.

Conversely, Quantum Computing~(QC) represents a promising approach in HEP~\cite{DiMeglio:2023nsa,Delgado:2022tpc,Rodrigo:2024say}, opening new paradigms for solving different types of challenging tasks where the quantum principles of entanglement and superposition can provide an advantage over classical methods. Recent examples include determining parton densities~\cite{Lamm:2019uyc,Perez-Salinas:2020nem,Chen:2025zeh}, simulating parton showers~\cite{Bepari:2021kwv,Bauer:2023ujy}, improving traking~\cite{Nicotra:2023rmn,Zlokapa:2019tkn}, jet clustering and evolution~\cite{Wei:2019rqy,deLejarza:2022bwc,Delgado:2022snu,Barata:2022wim}, and the efficient integration of multidimensional functions~\cite{Pyretzidis:2025stx,Williams:2025hza,deLejarza:2024scm,deLejarza:2024pgk,Cruz-Martinez:2023vgs,
deLejarza:2023qxk,Agliardi:2022ghn,Herbert:2021xgs}, among other topics. They include, in particular, the selection of the causal configurations of multiloop Feynman diagrams~\cite{Ramirez-Uribe:2021ubp,Clemente:2022nll,Ramirez-Uribe:2024wua}, which is equivalent to querying Directed Acyclic Graphs (DAGs) in graph theory.

A Grover's based quantum algorithm for searching DAG configurations was recenlty proposed using multicontrolled Toffoli gates (MCX)~\cite{Ramirez-Uribe:2024wua} as the most efficient quantum gates to tag cycles in the quantum oracle. While the algorithm can be extended to arbitrary complex topologies, the current technological limitations prevent this from being systematised further. In particular, the limited number of qubits available in current quantum simulators imposes the most significant constraint, and calls for  minimising quantum resources for further advances, which would also help mitigate noise in quantum hardware.
After identifying the DAG configurations and reconstructing the LTD representation, the next step is integration in the Euclidean space of the loop three-momenta. This second step has been achieved in Refs.~\cite{Pyretzidis:2025stx,deLejarza:2024scm,deLejarza:2024pgk,deLejarza:2023qxk}
with QC integration algorithms. In this paper, we present a novel quantum algorithm for the querying of causal structures, dubbed Minimun Clique-optimised quantum Algorithm (MCA), which is based on the Minimum Clique Partition (MCP)~\cite{CliqueCoverPartition} problem in graph theory. This represents a logical advancement, given the close analogy of causality in Feynman diagrams with graph theory. In particular, we focus on automating the process of constructing the oracle operator and optimising the transpiled quantum circuit of the proposed quantum algorithm.

This document is organized as follows: in Sec.~\ref{sec:LTD}, we briefly introduce the ideas of causality in LTD; in Sec.~\ref{sec:Quantum}, we present the principal components of Grover's based quantum algorithms and its relations to DAGs; in Sec.~\ref{sec:OptandAut}, we introduce the MCA quantum algorithm and analyse different strategies for optimisation and automation of the quantum querying oracle; in Sec.~\ref{sec:Transpilation}, 
we present a transpilation analysis of the MCA quantum algorithm, and contrast it with the results of the MCX quantum algorithm~\cite{Ramirez-Uribe:2024wua}; in Sec.~\ref{sec:Application}, we present the implementation of the proposed quantum algorithm to new topologies at four and five loops, demonstrating how efficiently these are tackled with our methodology and finally, in Sec.~\ref{sec:Conclusion} we present the conclusions and outline the perspectives of the work.

\section{Causality from the Loop-Tree Duality}
\label{sec:LTD}

The computation of high-precision theoretical predictions for observables in HEP relies on efficiently integrating scattering amplitudes at high perturbative orders in QFT. It is well known that scattering amplitudes exhibit different kinds of singularities and that theoretical predictions can be extracted only after the cancelation of these singularities. The methodology for removing the divergences from ultraviolet configurations at high energy depends on the renormalization procedure, while soft and collinear configurations require the construction of suitable counterterms. One of the potential bottlenecks is the proliferation of non-causal singularities in loop Feynman diagrams. These singularities of the integrand are non-physical artifacts that, while expected to cancel upon integration, give rise to numerical instabilities in the integrand. In this direction, LTD~\cite{Catani:2008xa, Bierenbaum:2010cy, Bierenbaum:2012th, Buchta:2014dfa, Buchta:2015wna, Hernandez-Pinto:2015ysa, Sborlini:2016gbr, Sborlini:2016hat, Tomboulis:2017rvd, Driencourt-Mangin:2017gop, Jurado:2017xut,Driencourt-Mangin:2019aix, Runkel:2019yrs, Aguilera-Verdugo:2019kbz, Runkel:2019zbm,Capatti:2019ypt, Driencourt-Mangin:2019yhu, Capatti:2019edf,Plenter:2020lop, Prisco:2020kyb,Verdugo:2020kzh,Soper:1999xk,snowmass2020,Aguilera-Verdugo:2020kzc,Ramirez-Uribe:2020hes,Imaz:2025buf,Sborlini:2021owe,Ramirez-Uribe:2024rjg,LTD:2024yrb,Aguilera-Verdugo:2020nrp,TorresBobadilla:2021dkq,TorresBobadilla:2021ivx,Rios-Sanchez:2024xtv,Aguilera-Verdugo:2021nrn} has made significant progress by enabling the manifest cancellation of such non-physical singularities directly at the integrand level. This advancement is achieved by exploiting the causal structure of scattering and vacuum amplitudes. In the following, we provide the basic ideas behind the LTD and its connection with DAG configurations in graph theory~\cite{Ramirez-Uribe:2021ubp,Clemente:2022nll,Ramirez-Uribe:2024wua}.

A multiloop Feynman diagram or scattering amplitude in the Feynman representation with $P$ external particles, $\{p_j\}_P$, and $n$ propagators is defined as an integral in the Minkowski space of $\Lambda$ loop  momenta, $\{\ell_s\}_\Lambda$, 
\begin{align}
{\cal A}_{\rm F}^{(\Lambda)} = \int_{\ell_1 \ldots \ell_\Lambda} {\cal N} (\{\ell_s\}_\Lambda, \{p_j\}_P) \prod_{i=1}^n G_{\rm F}(q_i)~,
\label{eq:AF}
\end{align}
where the momentum $q_i$ of each Feynman propagator, $G_{\rm F}(q_i)$, is a linear combination of loop momenta and external momenta. For vacuum amplitudes, the external momenta are absent. The integration measure in dimensional regularisation~\cite{Bollini:1972ui,tHooft:1972tcz} reads $\int_{\ell_s} = -\imath \mu^{4-d} \int {\rm d}^d \ell_s/(2\pi)^d$, with $d$ the number of spacetime dimensions and $\mu$ an arbitrary energy scale.
The integrand consists of the product of Feynman propagators and a numerator, ${\cal N}(\{\ell_s\}_\Lambda, \{p_j\}_P)$, which depends on the particles and interactions involved. 

A Feynman propagator is represented in a way that makes the poles explicit,
\begin{align}
    G_{\rm F}(q_i)=\frac{1}{\left(q_{i,0}+q_{i,0}^{(+)}\right)\left(q_{i,0}-q_{i,0}^{(+)}\right)} ~, 
\end{align}
with $q_{i,0}^{(+)}=\sqrt{\qb_i^{\, \, 2}+m_i^2-\imath 0}$ the on-shell energy, $q_{i,0}$ and $\qb_i$ the energy and spacial components of $q_i$, respectively, $m_i$ the mass of the propagating particle and $\imath 0$ the customary Feynman complex prescription. Feynman propagators encode in fact the quantum superposition of a positive and a negative energy mode between two interaction vertices, or equivalently the propagation of a particle in either directions. Therefore, a Feynman propagator can formally be written as the superposition of two quantum mechanical states,  
\begin{align}
    G_{\rm F}(q_i) \equiv \frac{1}{\sqrt{2}}\left(\vert 0 \rangle+\vert 1 \rangle\right) \, ,
\end{align}
with $\ket{0}$ denoting propagation in one direction and $\ket{1}$ in the oposite direction. 

A multiloop Feynman diagram in the Feynman representation can be considered as a quantum superposition of $2^n$ states, of which only a subset has a physical meaning. From a diagrammatic interpretation, diagrams in which a particle departs from any interaction vertex and never returns to it are physically meaningful and are associated with acyclic states. On the contrary, a diagram where a particle departs from an interaction vertex and returns to the initial point is interpreted as a particle traveling back in time, thus breaking causality, and is associated with cyclic states. A consequence of cyclic states is the raising of singular configurations related to non-physical processes, which are costly computationally and numerically unstable. 

Within the LTD framework the cyclic states related to the non-physical processes are explicitly absent. The LTD representation of any multiloop scattering amplitude is computed through the iterative evaluation of Cauchy's residue theorem. As a result, the multiloop scattering amplitude is expressed as a sum of nested residues. To explicitly cancel all the non-causal contributions, it is sufficient to sum together all the nested residues. After a convenient rearrangement, we achieve a causal LTD expression of the form
\beq
    \mathcal{A}_{\text{D}}^{(L)} = \int_{\boldsymbol{\ell}_1 \ldots \boldsymbol{\ell}_L} 
    \frac{1}{x_n} \sum_{\sigma  \in \Sigma} \frac{{\cal N}_{\sigma(i_1, \ldots, i_{n-L})}}{\lambda_{\sigma(i_1)}^{h_{\sigma(i_1)}} \cdots \lambda_{\sigma(i_{n-L})}^{h_{\sigma(i_{n-L})}}}
    + (\lambda_p^+ \leftrightarrow \lambda_p^-)~,
\label{eq:CausalRep}
\eeq
where $x_n = \prod_{i=1}^n 2\qon{i}$ and $h_{\sigma(i)}=\pm$. The LTD causal representation is expressed in terms of causal propagators of the form $1/\lambda_p^\pm$, with
\beq
    \lambda_p^\pm = \sum_{i\in p} \qon{i} \pm k_{p,0}~,
    \label{eq:CausalProp}
\eeq
where $p$ is a partition of the on-shell energies. We have from \Eq{eq:CausalProp} that each $\lambda_p^\pm$ corresponds to a kinematic configuration in which the propagators' momenta flows belonging to a partition $p$ are aligned in the same direction along a line that divides the Feynman diagram into two subamplitudes. The sign of $k_{p,0}$ determines whether $\lambda_p^+$ or $\lambda_p^-$ becomes singular once the propagators in the partition $p$ are set on shell.

Combinations of causal propagators with compatible momentum flows represent causal thresholds that are simultaneously possible, i.e., entangled causal thresholds. In \Eq{eq:CausalRep}, the set $\Sigma$ collects all the combinations of entagled causal thresholds. Each element in $\Sigma$ determines the momentum flow of all propagators in specific directions. Once the momentum flow of the propagators of each element has been fixed the LTD causal representation with the structure exhibited in \Eq{eq:CausalRep} can be established. A significant challenge to bootstrap the \Eq{eq:CausalRep} is the determination of the set $\Sigma$ by the identification of all internal configurations fulfilling causal conditions. As the topological complexity of a diagram increases, the number of possible causal propagators grows rapidly, as well as the number of combinatorial configurations of entangled causal propagators.


\section{Quantum algorithms for querying causality of multiloop graphs}
\label{sec:Quantum}
 
The aim of the quantum causal query algorithms presented in Refs.~\cite{Ramirez-Uribe:2021ubp,Ramirez-Uribe:2024wua}, relating causality of multiloop Feynman diagrams with DAGs, is the bootstrapping of the LTD causal representation in the most efficient way. Each term in the LTD causal representation corresponds to a configuration in which internal particles propagate along specific directions. Conversely, by fixing a given propagation pattern, one can reconstruct the associated LTD terms. These algorithms are based on analysing \textit{reduced} equivalent graphs built from vertices and edges~\cite{TorresBobadilla:2021ivx,Sborlini:2021owe}. An edge is defined as the union of multiple Feynman propagators connecting two interaction vertices. After the propagators are merged into edges, the loops associated to the reduced graph are called eloops. This reduction framework is well justified, as causal configurations are only those in which the momentum flows of all propagators belonging to each edge, are aligned in the same direction. In addition, considering eloops simplifies the analysis of causal conditions by reducing the number of acyclic configurations to identify.

A quantum query algorithm based on Grover's algorithm~\cite{Grover:1997fa} involves the following stages: $i)$ encoding the relevant information in qubits, where each qubit represents an edge; $ii)$ initializing all qubits representing the edges in the quantum circuit in a uniform superposition; $iii)$ designing an oracle operator that tags the acyclic states, which requires additional ancillary qubits; $iv)$ applying a diffusion operator to amplify the amplitude of the tagged states and $v)$ measuring the resulting quantum state, which, due to amplitude amplification, is most likely an acyclic state.

The distinguishing feature of a quantum query algorithm is the design of the oracle operator. In particular, the oracle operator in the quantum algorithm in Ref.~\cite{Ramirez-Uribe:2021ubp} considers binary clauses to compare adjacent edges, which are used to encode the causal conditions to be tested. Meanwhile, the oracle operator in the quantum algorithm MCX from Ref.~\cite{Ramirez-Uribe:2024wua}, directly encodes the causal conditions through multicontrolled Toffoli and XNOT gates. A multicontrolled Toffoli gate is a reversible quantum logic gate that generalizes the standard $3$-qubit Toffoli (CCNOT) gate to $n$ control qubits and one target qubit. A multicontrolled Toffoli gate flips the state of the target qubit if and only if all control qubits are in the $|1\rangle$ state. It is therefore the most natural choice for antitagging cycles, as it acts precisely when all edges in a cycle share the same orientation. This idea already simplifies the design of the oracle operator. However, the oracle operator in both algorithms from Refs.~\cite{Ramirez-Uribe:2021ubp,Ramirez-Uribe:2024wua} needs to be manually constructed, specifically, the selection of the minimum number of causal conditions and the establishment of an efficient order for implementing the quantum gates do not follow a particular methodology.

The MCA algorithm presented in this work encodes the causal conditions as proposed in the MCX algorithm, i.e. with multicontrolled Toffoli and XNOT gates, but regarding the oracle operator design, the MCA algorithm aims to achieve an optimal and automated oracle operator design by the application of a structured methodoloy, which efficiently reduces the ancillary quantum resources required and implements the causal conditions in a streamlined form. In order to optimise the quantum circuit, we focus on reducing the number of ancillary qubits required by exploiting the concept of Mutually Exclusive Clauses (MEC). This property allows information from different control qubits to be stored in a common target qubit. In the context of automating the MCA algorithm, and given the analogy with graph theory, we explore the use of graph features as a data structure to search for MECs, and also apply them to the efficient design of the oracle operator.

To evaluate the efficiency of the proposed MCA quantum algorithm, we analyse the number of qubits needed for the algorithm implemetation, the theoretical quantum circuit depth and the quantum circuit area~\cite{Ramirez-Uribe:2024wua}. The quantum circuit area is defined by the product of the transpiled quantum circuit depth and the number of qubits required in the transpilation. Before delving into the description of the MCA quantum algorithm, we review the concepts related to the Boolean construction of the causal clauses and describe the quantum circuit components.

\subsection{Boolean construction of eloop clauses}
\label{sec:boo_construction}

\begin{figure}[tb]
\centering
\includegraphics[width=\textwidth]{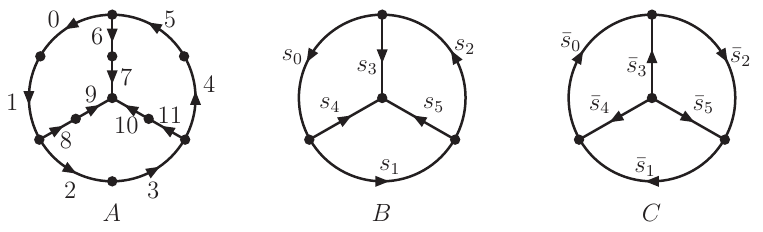}
\caption{Left ($A$): A three-eloop topology with two edges per set. Each set collects all the Feynman propagators that depend on the same linear combination of loop momenta. The edges are labeled considering the short hand notation $i \equiv e_i$ and the direction of the arrows represent the $\ket{1}$ states. External particles that may be attached to the vertices are not shown. Center ($B$) and Right ($C$): Boolean functions defined in \Eq{eq:s_tag} applied to the three-eloop topology in $A$.}
\label{fig:Sunrise_example}
\end{figure}

Causal configurations are identified by suppressing the amplitude of cyclic ones. Directed cyclic conditions are stored in eloop clauses, which are the logical conditions identifying a set of edges oriented in the same direction, thereby forming a closed cycle in a Feynman diagram. We introduce $z_j$ as the set of indices corresponding to edges that depend on the same linear combination of loop momenta. Given the set $z_j$ we define the following Boolean functions,
 \beq 
 s_j \equiv \bigwedge_{i \in z_j} e_i~, \qquad \text{and} \qquad  \bar{s}_j \equiv \bigwedge_{i \in z_j} \bar{e}_i,
\label{eq:s_tag}
 \eeq
where $e_i$ represents the state of edge $i$, the Boolean function $s_j$ is defined with the AND operator, $\wedge$, and $\bar{s}_j$ represents the Boolean function with the mirror states of the edges within the specified set, $\bar e_i = \neg e_i$ . The total number of edges constituting a multiloop topology is given by $n = \sum_{j} n_{s_j}$, where $n_{s_j}$ denotes the total number of edges in the set~$z_j$. For example, the Boolean functions associated to the three-eloop topology depicted in Fig.~\ref{fig:Sunrise_example}A are given by (see Fig.~\ref{fig:Sunrise_example}B)
\beq
\begin{aligned}
s_0 &= e_0 \wedge e_1~, & s_1 &= e_2 \wedge e_3~, & s_2 &= e_4 \wedge e_5~, \\
s_3 &= e_6 \wedge e_7~, & s_4 &= e_8 \wedge e_9~, & s_5 &= e_{10} \wedge e_{11}~.
\end{aligned}
\eeq
Fig.~\ref{fig:Sunrise_example}C represents the mirror Boolean functions. 

\begin{figure}[tb]
    \centering
    \includegraphics[width=\textwidth]{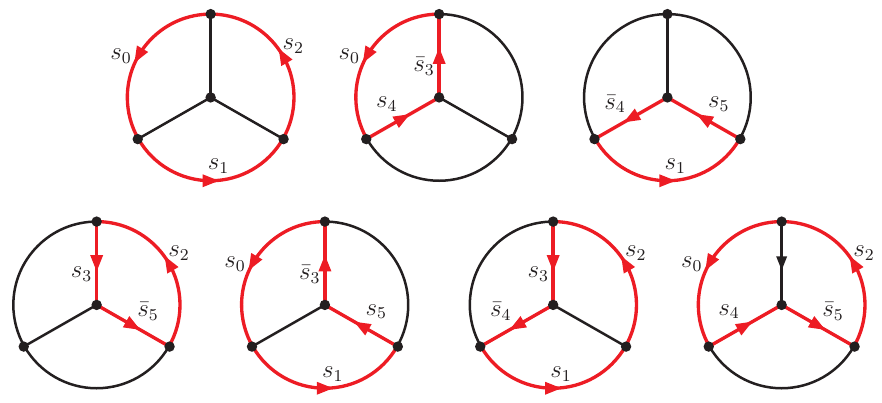}
    \caption{The eloop clauses from \Eq{eq:3_clauses}. The remaining seven eloop clauses are obtained by mirroring the edge states.}
    \label{fig:eloop_clauses}
\end{figure}

Relying on \Eq{eq:s_tag} enables to simplify the encoding of the eloop clauses, in such a way that they still depend on the number of sub-eloops but not longer on the number of edges. The maximal number of eloop clauses is determined by the total number of directed cyclic configurations, $m=2\ell$, where $\ell$ denotes the total number of sub-eloops. The eloop-clause conditions are given by
\beq 
c_k \equiv \bigwedge_{j\in S_k} s_j~, \qquad c_{\ell+k} \equiv \bar{c}_k \equiv \bigwedge_{j\in S_k} \bar s_j~, \qquad \text{with}\quad k=0,\cdots,\ell-1~, 
\label{eq:eloop_clauses}
\eeq
where capital $S_k$ denotes the set of edges generating the cyclic configuration contained in the sub-eloop $k$. Back to the three-eloop example in Fig.~\ref{fig:Sunrise_example}A, there are three sub-eloops involving three sets of edges and four sub-eloops involving four sets of edges, therefore the maximal number of eloop clauses is fourteen. Explicitly, the eloop clauses are 
\beq
\begin{aligned}
& c_0  = s_0 \wedge s_1 \wedge s_2~,  && c_1  = s_0 \wedge \bar{s}_3 \wedge s_4~, && c_2  = s_1 \wedge \bar{s}_4 \wedge s_5~, \\
& c_3  = s_2 \wedge s_3 \wedge \bar{s}_5~, && c_4  = s_0 \wedge s_1 \wedge \bar{s}_3 \wedge s_5~, && c_5 = s_1 \wedge s_2 \wedge s_3 \wedge \bar{s}_4~, \\
& c_6  = s_0 \wedge s_2 \wedge s_4 \wedge \bar{s}_5~, && c_{7+k} = \bar{c}_k~, &
\label{eq:3_clauses}
\end{aligned}
\eeq
with $k=0,\ldots,6$. A graphical illustration of these eloop clauses is  depicted in Fig.~\ref{fig:eloop_clauses}. Once the eloop clauses $c_k$ are established, the set of acyclic configurations is composed by those fullfiling the following Boolean condition
\beq
\mathcal{A} \equiv \bigwedge_{j} \neg c_j~,
\label{DAG_condition}
\eeq
suppressing the cyclic configurations. The representation of eloop clauses through Boolean functions is motivated from the fact that, at the quantum circuit level, the Boolean operator $\wedge$ is associated with a multicontrolled Toffoli gate~\cite{Ramirez-Uribe:2024wua}. 

\subsection{Quantum query algorithms}
\label{sec:Algorithms}

Grover's quantum algorithm aims to query a specific number of $r$ winning states over a total number of $N=2^n$ states, with $n$ the number of qubits generating all the possible states. Specifically, $n$ represents the number of edges in a Feynman diagram of interest, $N$ is the total number of possible configurations, and $r$ stands for the causal configurations or DAGs. The amplitude amplification is effective only under specific conditions. The mixing angle, $\theta = \arcsin(\sqrt{r/N})$, measures the ratio between the number of winning states and the total number of states. Grover's quantum algorithm is considered an appropriate approach if the mixing angle satisfies the condition given by $\theta\lesssim\pi/6~\left(r/N\lesssim1/4\right)$. 

The uniform superposition of the total of $N=2^n$ states is given by
\beq
\ket{e}= \frac{1}{\sqrt{N}} \sum_{x=0}^{N-1} \ket{x}~. 
\eeq
It can be understood as a vector collecting all the winning states $\ket{w}$, i.e. encoding all the causal solutions in a uniform superposition, and the orthogonal state $\ket{e_\perp}$, collecting the non-causal states
\beq
\ket{e} = \sin\theta \, \ket{w} + \cos \theta \, \ket{e_\perp}~.
\eeq
The winning and orthogonal uniform superpositions are given by
\beq
\ket{w}= \frac{1}{\sqrt{r}} \sum_{x\in w} \ket{x}~, \qquad 
\ket{e_\perp}= \frac{1}{\sqrt{N-r}} \sum_{x\notin w} \ket{x}~.
\eeq 
The oracle operator, denoted $U_w$, flips the state $\ket{x}$ if $x\in w$, $U_w \ket{x} = - \ket{x}$, and leaves it unchanged otherwise, $U_w \ket{x} = \ket{x}$ if $x\notin w$. The diffusion operator, $U_s = 2|e\rangle\langle e|- \boldsymbol{I}$, performs a reflection around the initial state $\ket{e}$, in order to amplify the probability of the winning states. According to the particular problem, the application of the oracle and diffusion operators may have to be applied more than once. The optimal number of iterations ($t$) is obtained when $\sin^2 \theta_t \sim 1$ with $\theta_t = (2t + 1)\theta$. Particularly, $t=1$ is sufficient when $\theta\sim\pi/6$ equivalent to $r/N\sim1/4$.

The multiloop topologies that we consider in the analysis are those included in the clasification scheme of multiloop topologies provided in Refs.~\cite{Verdugo:2020kzh,Ramirez-Uribe:2020hes,Ramirez-Uribe:2022sja} allowing to describe any scattering amplitude up to five loops. From a classical~\cite{Aguilera-Verdugo:2020kzc,Ramirez-Uribe:2020hes} and quantum~\cite{Ramirez-Uribe:2021ubp} approach, it has been shown that the number of causal configuratios is typically about one half of the total number of possible configurations, which is not optimal for amplitude amplification. In order to circumvent this limitation it is required to reduce the ratio between the number of configurations to query and the total number of possible configurations. The two strategies to achieve an optimal amplitude amplification are either to halve the number of causal configurations to query by tagging the state of one edge, or to increase the total number of possible states by introducing  additional qubits.
 
By tagging the orientation of one edge, we leverage the symmetry inherent to all diagrams, which satisfy the property that when a causal solution is provided, its mirrored state with all momentum flows reversed is also a causal solution. It is important to highlight that this property also reduces the number of Boolean clauses necessary for the construction of the oracle. If this modification is not sufficient to obtain a proper value of $r/N$, we increase the value of $N$ by adding an ancillary qubit. This adjustment is reserved as a last option, as increasing the number of qubits also increases the quantum circuit depth.

Recaping the example of the three-eloop topology in Fig.~\ref{fig:Sunrise_example}A, we have that $r/N\sim1/2$. So, we tag the state of one edge, adopting as a convention the edge $e_0$. Reformulating and relabeling the indices of the eloop clauses shown in Eq.~(\ref{eq:3_clauses}) results in a reduction from fourteen to ten clauses: 
\beq
\begin{aligned}
c_0  &= s_0 \wedge s_1 \wedge s_2~, & c_1  &= s_0 \wedge \bar{s}_3 \wedge s_4~, & c_2  &= s_1 \wedge \bar{s}_4 \wedge s_5~, \\
c_3  &= s_2 \wedge s_3 \wedge \bar{s}_5~, & c_4  &= s_0 \wedge s_1 \wedge \bar{s}_3 \wedge s_5~, & c_5  &= s_1 \wedge s_2 \wedge s_3 \wedge \bar{s}_4~, \\
c_6  &= s_0 \wedge s_2 \wedge s_4 \wedge \bar{s}_5~, & c_7  &=\bar{c}_2~, \quad~ c_8 = \bar{c}_3~,& c_{9} &= \bar{c}_5 ~. & 
\end{aligned}
\label{3_clauses_reduced}
\eeq

\begin{figure}[t!]
    \centering \includegraphics[width=\textwidth]{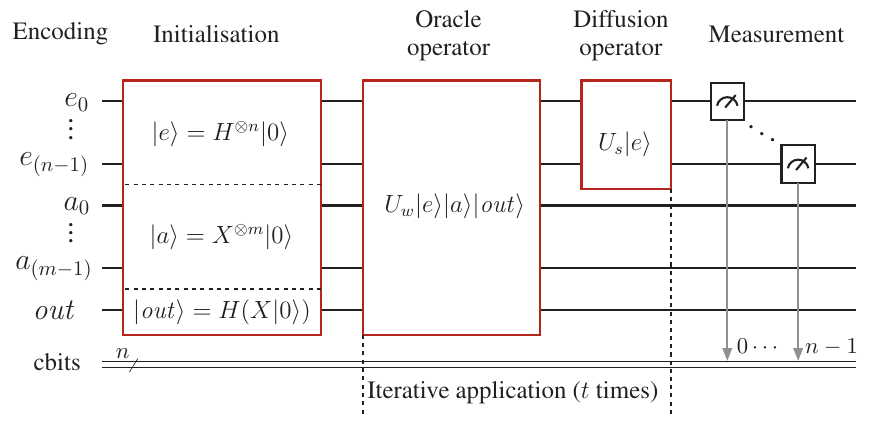}
    \caption{Quantum circuit model of a quantum query algorithm. The scheme of the procedure is composed by five stages: encoding, the registers initialisation, the oracle operator, the diffusion operator and the measurement.}
    \label{fig:flujo_circuito}
\end{figure}

At a quantum circuit level the quantum circuit model to implement a quantum query algorithm based in Grover's quantum algorithm is illustrated in Fig.~\ref{fig:flujo_circuito}. The first step is to encode the required information, in this case the quantum algorithm needs three different classes of qubit registers. The edges defining a multiloop topology are encoded in the first register denoted by $\ket{e}$. The second register stores the causal eloop clauses in ancillary qubits, and is denoted by $\ket{a}$. The last register encodes the oracle marker, $\ket{\text{out}}$, which distinguishes whether the graph configuration corresponds to a causal solution or not. 

Once the three registers have been incorporated into the quantum circuit, the next step is to initialise them. The $\ket{e}$ register is initialised in a uniform superposition by applying Hadamard gates to the $n$ qubits characterizing the edges, $\ket{e} = H^{\otimes n}\ket{0}$; in the case of the $\ket{a}$ register, the $m$ ancillary qubits corresponding to the $m$ eloop clauses are initialised to the $\ket{1}$ state through the application of XNOT gates, $\ket{a} = X^{\otimes m}\ket{0}$; and the oracle marker is initialised to the Bell state $\ket{-}$ by $\ket{\text{out}} = \ket{-1} = H(X\ket{0})$.

After the initialisation of all registers has been completed, the oracle operator encoding all eloop clauses is applied through the phase kick-back property, 
 \beq
 U_{w} \ket{e}\ket{a}\ket{\text{out}} = (-1)^{f(a,e)}\ket{e}\ket{a}\ket{\text{out}}~,
 \eeq
where $f(a,e) = (\wedge_i a_i) \wedge e_0$, and $e_0$ is the edge whose state is tagged in a fixed orentation. Before moving forward to the probability amplification by the diffusion operator, the oracle operator is applied in reverse order to restore each qubit to its initial state, with the exception of $\ket{\text{out}}$. Once this point is reached, the diffusion operator, $U_s$, is applied to the $\ket{e}$ register. In this work, the diffusion operator used is the one described in the \texttt{PennyLane} documentation~\cite{Bergholm:2018cyq}.

Up to this point, there are no structural differences between the MCX and MCA quantum algorithms. However, the MCA algortihm takes one step further by providing a well-defined methodological approach for designing an oracle operator that minimise the number of ancillary qubits required and optimise the order in which quantum gates are implemented through graph theory techniques. The following section shows in detail the design of the oracle operator, taking into account the impact of the number of eloop clauses in the required computational resources.

\section{The Minimum Clique-optimised quantum algorithm}
\label{sec:OptandAut}

The effectiveness of quantum query algorithms based on amplitude amplification critically relies  on the oracle operator design. The eloop clauses required to construct the oracle operator are established according to the diagram complexity scaling polynomially. Therefore, it is important to have in mind that a larger number of clauses and the implementation in an arbitrary order of the quantum gates encoding the eloop clauses implies a higher resource consumption either in the number of qubits, the number of quantum gates, or the quantum depth of the quantum circuit. 

Regarding the eloop clauses, Sec.~\ref{sec:boo_construction} presents the logic to define the eloop clauses; Sec.~\ref{sec:Algorithms} considers the fact that given a causal solution its mirror state is also a causal solution to reduce by half the number of causal states to query. In this section we present the MCA approach to optimise the number of eloop clauses and quantum gates needed, as well as an automated procedure to apply them in the optimal order. The MCA quantum algorithm is characterized by the use of fundamental graph theory concepts to process the graph structure more efficiently, with the aim of having a positive impact on reducing the number of qubits required in the eloop-clause quantum register~$\ket{a}$ and establishing an appropriate order for implementing the required quantum gates. Furthermore, this reduction results in lowering the theorical quantum depth of the quantum circuit, which has a direct impact on reducing the expected quantum noise in quantum hardware.

\subsection{Graphs as data structures and search of clique}

Graph theory has become an important mathematical tool in a wide variety of applications. Particularly in computer science, graphs have emerged as a fundamental data structure~\cite{preiss1999data} for modelling relationships between pairs of objects. One of the most interesting features of this data structure is its inherent ability to represent multiple relationships between objects, which categorizes it as a non-linear data structure. This nonlinearity has enable the development of algorithms capable of solving a diverse range of problems~\cite{6005872}.

\begin{figure}[htb]
\centering
\includegraphics[width=0.5\textwidth]{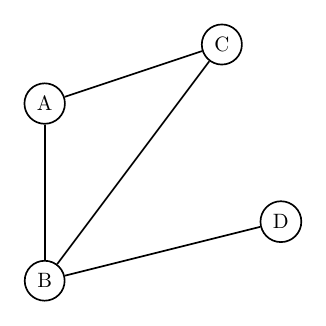}
\caption{Graph consisting of vertices A, B, C and D, with a clique composed by the set of vertices $\{A, B, C \}$.}
\label{fig:graph}
\end{figure}

A fundamental concept used to design and automate the MCA quantum algorithm is the identification of cliques in an undirected graph. A clique, also known as a complete subgraph, is defined as a subset of vertices from a specific graph in which each pair of vertices is directly connected by an edge. For example, the graph depicted in Fig.~\ref{fig:graph} contains a clique composed by the set of vertices $\{A, B, C \}$. 
In particular, we focus on determining the minimum number of cliques that generate a graph, a problem known as Minimum Clique Partition (MCP)~\cite{CliqueCoverPartition}. To clarify the idea behind the MCP problem, we recall the graph shown in Fig.~\ref{fig:graph}, which MCP solution is given by the configuration of the clique generated by the set of vertices $\{A, B, C\}$ and the clique formed solely by the vertex~$D$. 

Regarding Feynman diagrams, a clique can be understood as a subset of interaction vertices that form non-causal configurations independently of the rest of the diagram. The advantage of associating eloop clauses with a subgroup of vertices defining a graph, and characterizing the corresponding edges with the desired criteria, is that an eloop clause optimisation problem can be treated as a MCP problem of the corresponding graph. It is also important to note that the MCP problem is NP-hard~\cite{karp2010reducibility}, which means that a fully general solution to this problem is beyond the scope of this work. Nevertheless, we provide an approximate solution that has proven to be suitable for our purposes.

\subsection{Ancillary qubits optimisation}
\label{sec:Optimisation}
The idea behind the MCA quantum algorithm to optimise the number of ancillary qubits for encoding of the eloop clauses is based on the concept of “mutually exclusive clauses", i.e., clauses that cannot be satisfied simultaneously for a given diagram configuration. An example of mutually exclusive clauses is the pair of clauses $c_2$ and $c_7$ of Eq.~(\ref{3_clauses_reduced}), declaring one of them true automatically implies that the other clause is false. It is important to note that by associating the concept of mutually exclusive clauses with a truth table, it is possible to establish the equivalence of mutually exclusive clauses with the Boolean XOR operator~$(\veebar)$. 

A useful relationship amoung the operators OR, AND and XOR is given by
\beq 
c_i\vee c_j = \left(c_i\veebar c_j\right) \vee \left(c_i\wedge c_j\right)~,  \qquad  \forall \, i \ne j~.
\label{general_boolean_property}
\eeq 
If $c_i \wedge c_j = \emptyset$, it is said that $c_i$ and $c_j$ ($i \neq j$) are mutually exclusive clauses and Eq.~(\ref{general_boolean_property}) simplifies to
\beq 
c_i\vee c_j = c_i\veebar c_j, \qquad \forall \, i \neq j~.
\label{eq:mutual_condition}
\eeq 
At the quantum circuit level, the condition given by \Eq{eq:mutual_condition} indicates the feasibility to store the information of both clauses in a single target qubit. The identification of mutually exclusive clauses allows a significant reduction in the number of required ancillary qubits.

\begin{algorithm}
\caption{\texttt{MutualAuxMatrix}}
\label{alg:MutualAuxMatrix}
\begin{algorithmic}[1]
    \STATE \textbf{Input:} $c_1, c_2, \dots, c_n$ \# All clauses
    \STATE \textbf{Output:} \texttt{aux\_matrix} \# $n \times n$ matrix
    \STATE num\_clauses $\gets$ $n$
    \STATE \texttt{aux\_matrix} $\gets$ empty matrix of size $n \times n$
    \FOR{$i = 1$ \TO num\_clauses}
        \FOR{$j = i$ \TO num\_clauses}
            \IF{$c_i \vee c_j = c_i \veebar c_j$}
                \STATE \texttt{aux\_matrix}[$i,j$] $\gets 1$
            \ELSE
                \STATE \texttt{aux\_matrix}[$i,j$] $\gets 0$
            \ENDIF
        \ENDFOR
    \ENDFOR
    \RETURN \texttt{aux\_matrix}
\end{algorithmic}\label{alg:uno}
\end{algorithm}

The problem of optimising the number of ancillary qubits can be treated as the search of the smallest number of  mutually exclusive subset clauses. The relation between clauses is illustrated through a graph, where a mutually exclusive clause set is represented with a clique.  In order to address the problem in the realm of graph theory, we construct the adjacency matrix, which describes the relation between eloop clauses. The \textbf{Algorithm}~\ref{alg:uno}, called \texttt{MutualAuxMatrix}, constructs a matrix whose entries are composed by zeros and ones; if the entry is one, the pair of clauses satisfies Eq.~(\ref{eq:mutual_condition}), otherwise the entry is zero. 

\begin{figure}[b]
    \includegraphics[scale = 0.46]{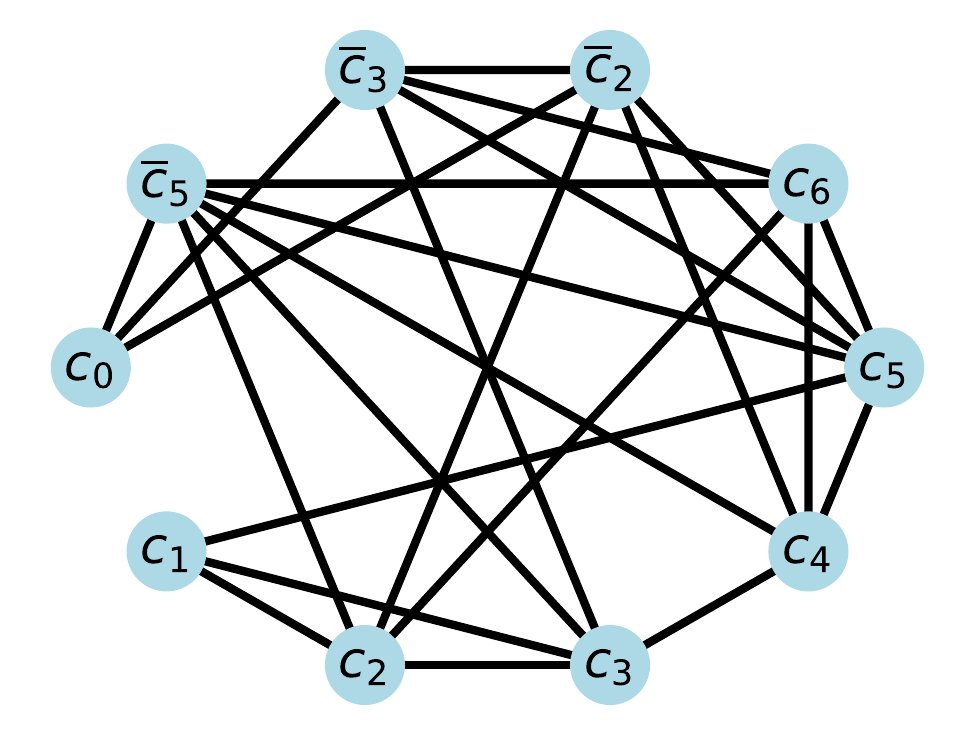}
    \includegraphics[scale = 0.51]{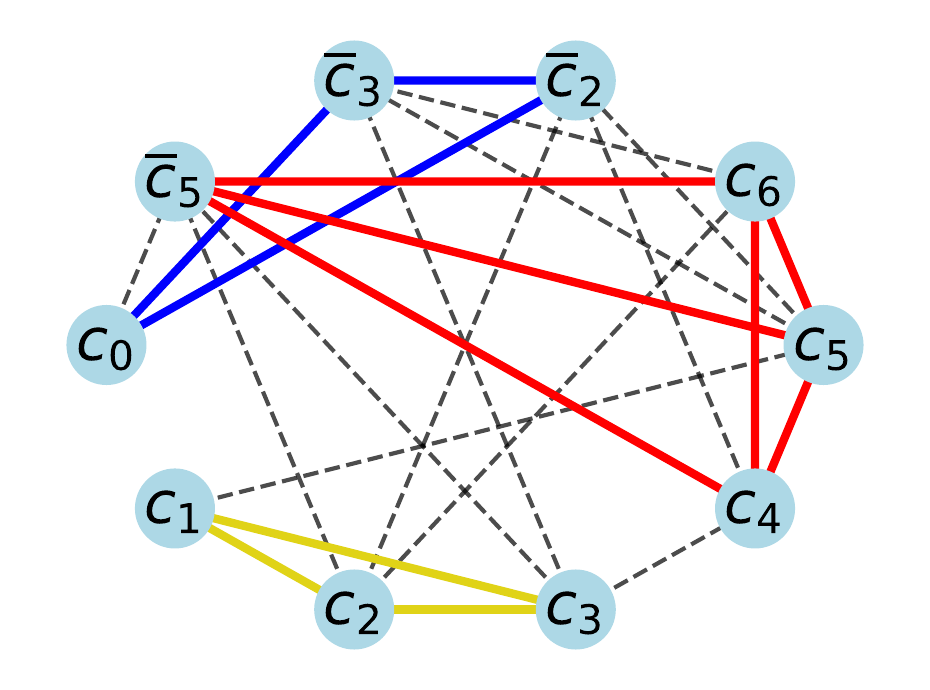}
    \caption{
    Graph representing the adjacency matrix of mutually exclusive clauses of the three-eloop topology in Fig.~\ref{fig:Sunrise_example}A, generated with the \texttt{MutualAuxMatrix} algorithm (left), and corresponding \texttt{MAUXc}$^{(3,12)}$ obtained with the \texttt{GraphConditionCombination} algorithm (right). Colours indicate the different cliques.}
    \label{fig:graph_mutual_exclusive_2}   
\end{figure}

The graph associated to the generated adjacency matrix is a practical tool to visualise the existence of multiple combinations of mutually exclusive clauses. Recalling the three-eloop topology in Fig.~\ref{fig:Sunrise_example}A, the Fig.~\ref{fig:graph_mutual_exclusive_2}~(left) shows the corresponding graph of the adjacency matrix of mutually exclusive clauses representing the eloop clauses in Eq.~(\ref{3_clauses_reduced}). 

In order to find the minimum number of cliques in a graph, the \textbf{Algorithm}~\ref{alg:dos} called \texttt{GraphConditionCombination} has been developed. This algortihm uses the function \texttt{find\_cliques}\footnote{This function is based on the algorithm described in Ref.~\cite{hagberg2008exploring}} from \texttt{Networkx} library \cite{10.1145/362342.362367} which obtains all cliques in the provided adjacency matrix. The key is to identify the \textit{maximum clique}, the clique containing the highest number of clauses. Once the maximum clique has been selected, the corresponding combination of clauses is stored and the associated nodes are removed from the graph. This process is equivalent to delete the rows and columns corresponding to each clause in the adjacency matrix, and is repeated until all the nodes of the graph are removed. The output consists of a set of mutually exclusive subsets of clauses; in terms of the quantum circuit it defines a set of Mutual Auxiliary clauses, \texttt{MAUXc}$^{(\Lambda,e)}$, where the superidex $(\Lambda,e)$ indicates the number of eloops and the number of edges respectively. Specifically, the number of sets in \texttt{MAUXc}$^{(\Lambda,e)}$ establishes the number of ancillary qubits required and the eloop clauses belonging to the same set indicates that they are stored in the same ancillary qubit.

\begin{algorithm}
\caption{\texttt{GraphConditionCombination}}
\label{alg:Graph_condition_combination}
\begin{algorithmic}[1]
    \STATE \textbf{Input:} \texttt{conditional\_graph} \# A graph where we search for combinations based on a condition
    \STATE \textbf{Output:} \texttt{clauses\_combination} \# A list of maximal subgroups of cliques in the graph
    \STATE \texttt{clauses\_combination} $\gets$ empty list
    \WHILE{conditional\_graph has nodes}
        \STATE \texttt{max\_clique} $\gets$ the largest clique in \texttt{conditional\_graph}
        \STATE append \texttt{max\_clique} to \texttt{clauses\_combination}
        \STATE remove nodes in \texttt{max\_clique} from \texttt{conditional\_graph}
    \ENDWHILE
    \RETURN \texttt{clauses\_combination}
\end{algorithmic}\label{alg:dos}
\end{algorithm}
It is important to mention that the \texttt{GraphConditionCombination} algorithm depends only on the structure of the graph, therefore, it becomes an important tool to automate the design of the oracle operator.

The application of \textbf{Algorithm~2} to the adjacency matrix associated to Fig.~\ref{fig:graph_mutual_exclusive_2}~(left) yields to the cliques shown in Fig.~\ref{fig:graph_mutual_exclusive_2}~(right), corresponding to the following sets:
\beq 
\texttt{MAUXc}^{(3,12)} = \big\{\{c_4, c_5, \bar{c}_5, c_6\}, \{c_0, \bar{c}_2, \bar{c}_3\}, \{c_1, c_2, c_3\}\big\}~.
\label{eq:MAUXc_3e12}
\eeq
In this example, the number of required ancillary qubits is three, each one storing the eloop clauses corresponding to each set in \texttt{MAUXc}$^{(3,12)}$. The three ancillary qubits required represent a reduction of $\mathcal{O}(57\%)$ in the number of ancillary qubits compared to the seven ancillary qubits required with the MCX algorithm. It is important to highlight that the advantage of encoding eloop clauses through multicontrolled Toffoli gates and XNOT gates is that the result obtained in \Eq{eq:MAUXc_3e12} remains unchanged even if the number of propagators per set increases. 

\subsection{Oracle design automation}

The oracle operator is composed by the eloop clauses, which are implemented through multicontrolled Toffoli gates and XNOT gates, the XNOT gates play the role of generating the states $\bar{s}_j$. Given that the order in which the gates are implemented in the quantum circuit directly impacts the quantum circuit depth, in this section we present an automated optimisation process to reduce the quantum depth by defining an optimal order to apply the eloop clauses.

\begin{figure}
    \centering
    \includegraphics[scale = 0.45]{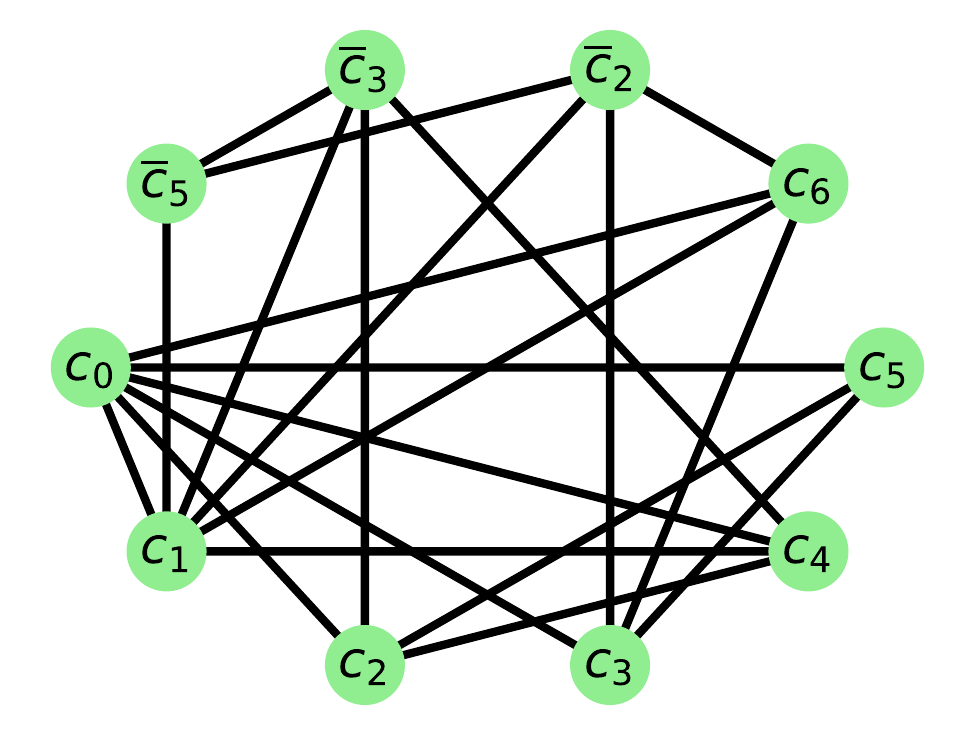}
    \includegraphics[scale = 0.48]{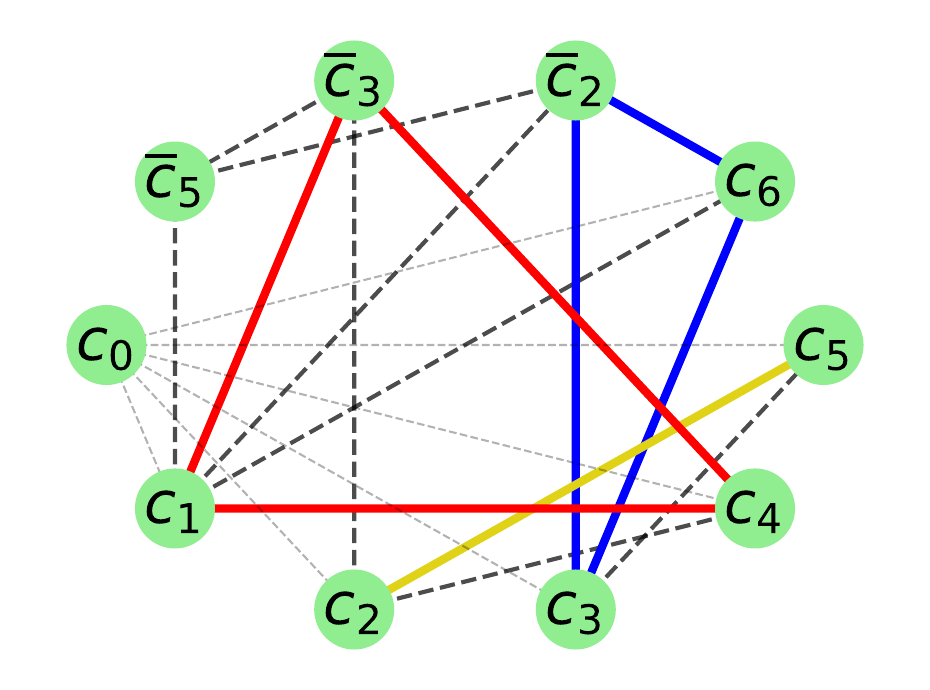}
    \caption{Graph representing the adjacency matrix of mutually compatible clauses for the three-eloop topology in Fig.~\ref{fig:Sunrise_example}A, generated with the \texttt{MutualClausesMatrix} algorithm (left), and corresponding optimised grouping of eloop clauses obtained with the \texttt{GraphConditionCombination} algorithm (right).}
    \label{graph_clauses_depth}
\end{figure}

The central idea is to group the eloop clauses from \texttt{MAUXc}$^{(\Lambda,e)}$ into sets, and then optimise the order in which the sets are implemented. Starting with the grouping of eloop clauses, we take advantage of the \texttt{MutualAuxMatrix} algorithm (Algorithm~\ref{alg:uno}) grouping structure to generate the algorithm called \texttt{MutualClausesMatrix} (Algorithm~\ref{alg:tres}) by modifying the clustering condition. The \texttt{MutualClausesMatrix} algorithm applied to the eloop clauses in \Eq{3_clauses_reduced} generates an adjacency matrix of mutually compatible clauses by establishing three clustering conditions. The first condition is to select the clauses that can be applied in the same time step, i.e., those clauses that have no common edges. The second condition considers the remaining eloop clauses of the diagram which match at least one state $s_k$ or $\bar{s}_k$. Explicitly, if $s_k\in c_i$ and $\bar{s}_k \in c_j$, then
\beq 
s_k \in  c_i \cap \bar{c}_j.
\label{eq:oracle_condition}
\eeq
At quantum circuit level, the condition in \Eq{eq:oracle_condition} allows us to apply a single column of XNOT gates, which prepares the corresponding states for all the clauses belonging to the set, followed by applying the corresponding multicontrolled Toffoli gates. The last condition takes as an independent set the eloop clause corresponding to the outer edges of the diagram.

The advantage of including this procedure in the oracle design is that, with the first condition we enable multiple eloop clauses to be applied in a single time step, with the second condition we reduce the number of XNOT gates by preventing additional contributions to the quantum depth through the application of a single column of XNOT gates, and with the last condition we take advantage that preparing the states associated with the eloop clause of the outer edges does not require XNOT gates because we always tag one of its edges in an specific state.

\begin{algorithm}
\caption{\texttt{MutualClauseMatrix}}
\label{alg:Graph_condition_combination}
\begin{algorithmic}[1]
    \STATE \textbf{Input:} $c_1, c_2, \dots, c_n$ \# All clauses
    \STATE \textbf{Output:} \texttt{com\_matrix} \# $(n-1) \times (n-1)$ matrix
    \STATE num\_clauses $\gets$ $n$
    \STATE ext\_clause $\gets$ $m$
    \STATE \texttt{com\_matrix} $\gets$ empty matrix of size $n \times n$
    \FOR{$i = 1$ \TO num\_clauses}
        \FOR{$j = i$ \TO num\_clauses}
            \IF{$c_i \vee c_j \neq c_i \veebar c_j$ ~\AND $i\neq j$ } 
                \STATE \texttt{com\_matrix}[$i,j$] $\gets 1$
            \ELSE
                \STATE \texttt{com\_matrix}[$i,j$] $\gets 0$
            \ENDIF
        \ENDFOR
    \ENDFOR
    \STATE Remove $m$-th row and $m$-th column from \texttt{com\_matrix}
    \RETURN \texttt{com\_matrix}
\end{algorithmic}\label{alg:tres}
\end{algorithm}

After the adjacency matrix of mutually compatible clauses is determined we continue to optimise the clustering of the compatible clauses, i.e. we find the minimum number of cliques. The optimisation is achieved by applying the \texttt{GraphConditionCombination} algorithm (Algorithm~\ref{alg:dos}) to the adjacency matrix of mutually compatible clauses. Specifically, it reduces the number of grouping blocks of clauses used to determine the arrangement of the gates. Then we add the external clause as its own subset in the set of eloop clauses.

Returning to the three-eloop example in Fig.~\ref{fig:Sunrise_example}A, the Fig.~\ref{graph_clauses_depth}~(left) shows a representative graph of the adjacency matrix of mutually compatible clauses obtained by the \texttt{MutualClausesMatrix} algorithm applied to the eloop clauses in Eq.~(\ref{3_clauses_reduced}). The application of the \texttt{GraphCondition\-Combination} algorithm to this adjacency matrix provides the graph shown in Fig.~\ref{graph_clauses_depth}~(right)
corresponding to the optimised sets of eloop clauses (\texttt{MUTc}$^{(\Lambda,e)}$):
\beq 
\texttt{MUTc}^{(3,12)} = \big\{ \{c_0\}, \{c_1, c_4, \bar{c}_3\}, \{ c_3, c_6,\bar{c}_2\}, \{c_2, c_5\}, \{\bar{c}_5\}\big\}~. 
\label{eq:MUTc_3e12}
\eeq
The coloured lines in Fig.~\ref{graph_clauses_depth}~(right) represent the cliques in \texttt{MUTc}$^{(3,12)}$ given by the first and second condition; the lightest dashed lines are given by the last condition which selects the external eloop clause, $\{c_0\}$, as an independent set. Once the \texttt{MUTc}$^{(3,12)}$ set has been obtained, the oracle operator is implemented by assembling the clause blocks in the order specified in \Eq{eq:MUTc_3e12}. In this case, the theoretical quantum circuit depth is twenty-nine.

\begin{figure}
    \centering
    \includegraphics[width=\textwidth]{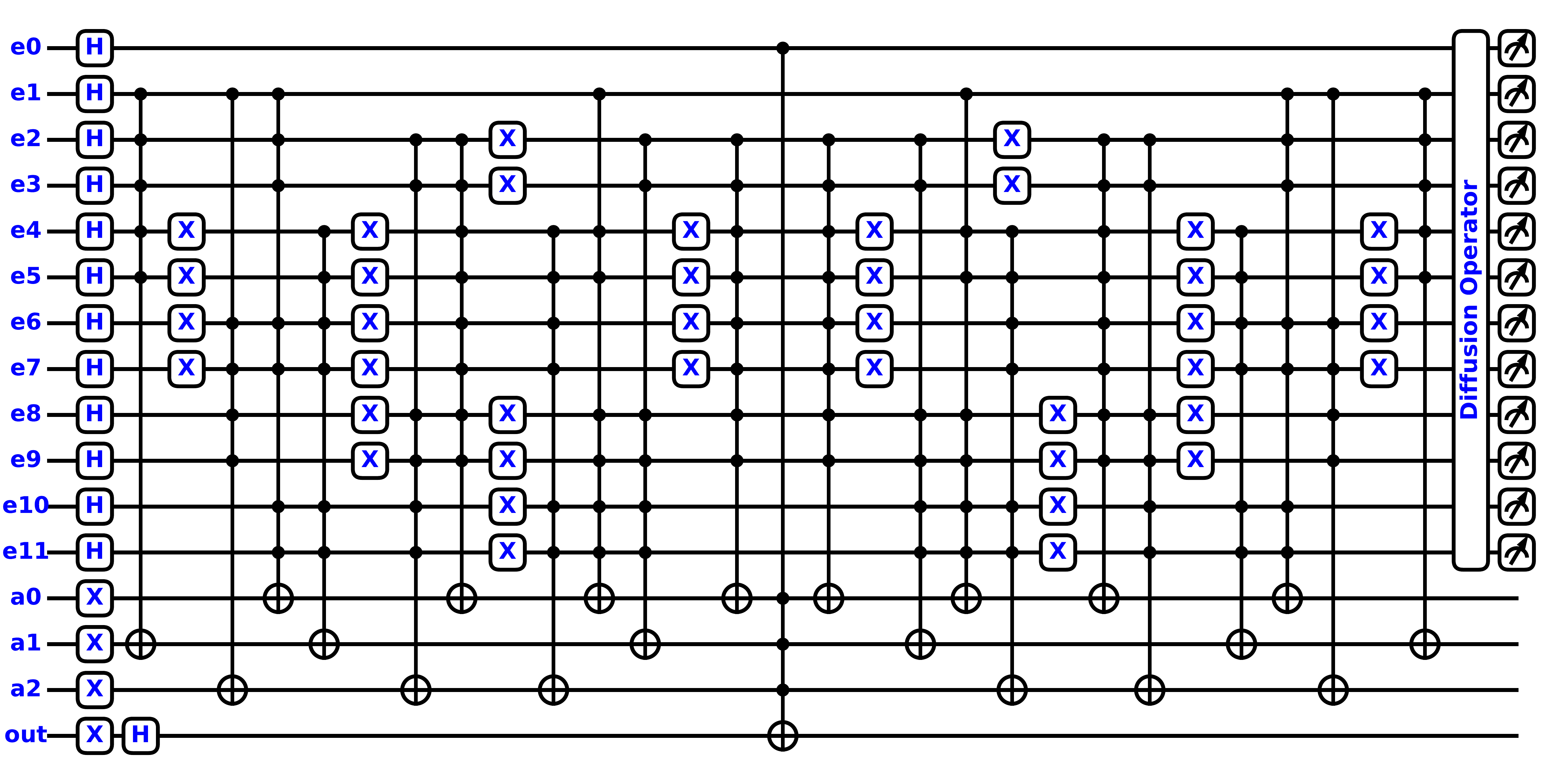}
    \caption{Quantum circuit associated for the three-eloop topology with two edges per line shown in Fig.~\ref{fig:Sunrise_example}A.}
    \label{fig:circuit_e12}
\end{figure}

It is important to highlight that the order of the sets contained in \texttt{MUTc}$^{(3,12)}$ is not unique, therefore, the quantum circuit depth depends on the order in which the sets are implemented. This scenario presents notable similarities to the process of identifying the optimal set of hyperparameters to maximise or minimise a specific metric for Machine Learning models. Specifically, we minimise the quantum circuit depth by generating an optimally ordered \texttt{MUTc}$^{(\Lambda,e)}$ set (\texttt{OMUTc}$^{(\Lambda,e)}$) applying \texttt{Optuna}~\cite{optuna}, a library widely used for hyperparameter optimisation of Large Deep Learning models. \texttt{Optuna} is based in the Tree-structured Parzen Estimator algorithm~(TPE)~\cite{TPEalgorthm}, which is a Bayesian optimisation algorithm that trains two Gaussian Mixture Models, one to the set of parameter values associated with the best objective values which is commonly denoted by $l(x)$, and another to the remaining parameter values, denoted by $g(x)$. From these two models it chooses the parameter value~$x$ that maximises the ratio $l(x)/g(x)$ and like a tree, performs a cut-off at this value of $x$ which provides the set of hyperparameter values to evaluate.

Recapping the three-eloop example in Fig.~\ref{fig:Sunrise_example}A, we apply \texttt{Optuna} to the set $\texttt{MUTc}^{(3,12)}$ in \Eq{eq:MUTc_3e12} obtaining
\beq 
\texttt{OMUTc}^{(3,12)} = \big\{ \{c_0\}, \{c_1, c_4,\bar{c}_3 \},\{c_2, c_5\},  \{c_3, c_6,\bar{c}_2 \} ,\{\bar{c}_5\}\big\}~. 
\label{eq:OMUTc_3e12}
\eeq
The implementation of clause blocks in the oracle operator in the optimal order allows us to reduce the theoretical quantum circuit depth from twenty-nine to twenty-three. It is important to note that obtaining \texttt{MUTc$^{(3,12)}$} must precede the application of \texttt{Optuna}, as it greatly reduces the space of posible permutations. The quantum circuit for the three-eloop topology in Fig.~\ref{fig:Sunrise_example}A is shown in Fig.~\ref{fig:circuit_e12}.

Gathering the procedure described in Section~\ref{sec:Quantum} and Section~\ref{sec:OptandAut} provides the complete methodology of the MCA quantum algorithm. Specifically, this section has presented the procedure that characterised the MCA quantum algorithm, with the three-eloop topology in Fig.~\ref{fig:Sunrise_example}A as an illustrative example. The sets contained in \texttt{MAUXc}$^{(3,12)}$ in Eq.~(\ref{eq:MAUXc_3e12}) represent the smallest set of eloop clauses for the most general three-eloop topology, i.e. with an arbitrary number of edges per set. Regarding the structure of the oracle operator \texttt{MAUXc}$^{(3,12)}$ given in \Eq{eq:MAUXc_3e12} provides the sets of mutually exclusive clauses and thus the number of ancillary qubits, the \texttt{MUTc}$^{(3,12)}$ sets given in \Eq{eq:MUTc_3e12} provide the most convenient clustering, whereas the sequence of sets in \texttt{OMUTc}$^{(3,12)}$ shown in \Eq{eq:OMUTc_3e12} introduces the optimal ordering for the implementation of the eloop clauses.

\section{Transpilation behaviour}
\label{sec:Transpilation}

\begin{figure}[t]
    \centering
    \includegraphics[width=\textwidth]
    {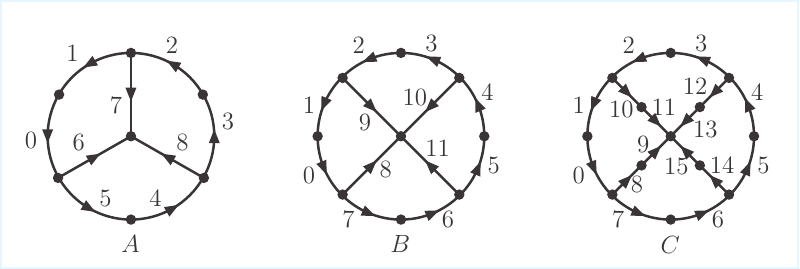} 
    \caption{From left to right: three-eloop topology with nine edges (A), four eloops with a four-point contact interaction involving twelve (B) and sixteen (C) edges.}
    \label{fig:3e9-4e12}
\end{figure}

It is important to note that the MCA quantum algorithm's advantage lies in reducing the number of ancillary qubits and theoretical quantum depth compared to MCX. In this section, we extend the  compararison of the MCX and MCA algorithms from the three-eloop topology in Fig.~\ref{fig:Sunrise_example}A to the eloop topologies in Fig.~\ref{fig:3e9-4e12}, which includes a three-eloop topology with nine edges, and four-eloop topologies with twelve and sixteen edges. 

Table~\ref{tab:comp_diag} summarises the quantum resources required to implement the MCX and MCA algoritms for these multiloop topologies. Regarding the required number of qubits, the fourth and fifth columns in Table~\ref{tab:comp_diag} exhibit a larger reduction for topologies of higher complexity, meaning MCA offers greater advantages for more intricate diagrams. In terms of the theoretical quantum circuit depth \cite{quantumstackexchange2020}, the sixth column in Table~\ref{tab:comp_diag} shows a similar behaviour, having a wider range of improvement in the more complex topologies. 

\begin{table}[t]
\centering
\begin{tabular}{lcccccc} \hline
    \textbf{Fig.} & \textbf{eloops (edges)} & $|e\rangle$ & $|a \rangle$ & \textbf{Total Qubits} & \parbox{2.5cm}{\textbf{Quantum Depth}}  & \textbf{Total states}\\ \hline
    \ref{fig:3e9-4e12}A & three (9) & 9 & $2\mid 4$ & $12\mid 14$ & $15 \mid 17$ & $512$ \\  
    \ref{fig:Sunrise_example}A & three (12) & 12 & $3\mid 7$ & $16\mid 21$ & $ 23 \mid 31$  & $8192$\\  
    \ref{fig:3e9-4e12}B & four$^{(c)}$ (12) & 12 & $4\mid 5$ & $17\mid 18$ & $15 \mid 15$ & $4096$\\  
    \ref{fig:3e9-4e12}C & four$^{(c)}$ (16) & $16+1$ & $6\mid 13$ &  $24\mid 31$ & $ 39\mid$ $45$ & $131072$ \\  \hline
    \end{tabular}
\caption{Quantum resources required and theoretical quantum circuit depth of the quantum algorithms used to analyse the three-eloop topology with nine and twelve edges, and the four-eloop topology with twelve and sixteen edges. The first number in the fourth, fifth and sixth columns are from the MCA quantum algorithm, whereas the second number corresponds to MCX quantum algorithm. The
total number of qubits includes the oracle’s marker. An ancillary qubit in the $\ket{e}$ register is necessary for the four-eloop topology with sixteen edges to achieve the optimal amplitude amplification.}
\label{tab:comp_diag}
\end{table}

\begin{figure}[b] 
{\includegraphics[width=.49\linewidth]{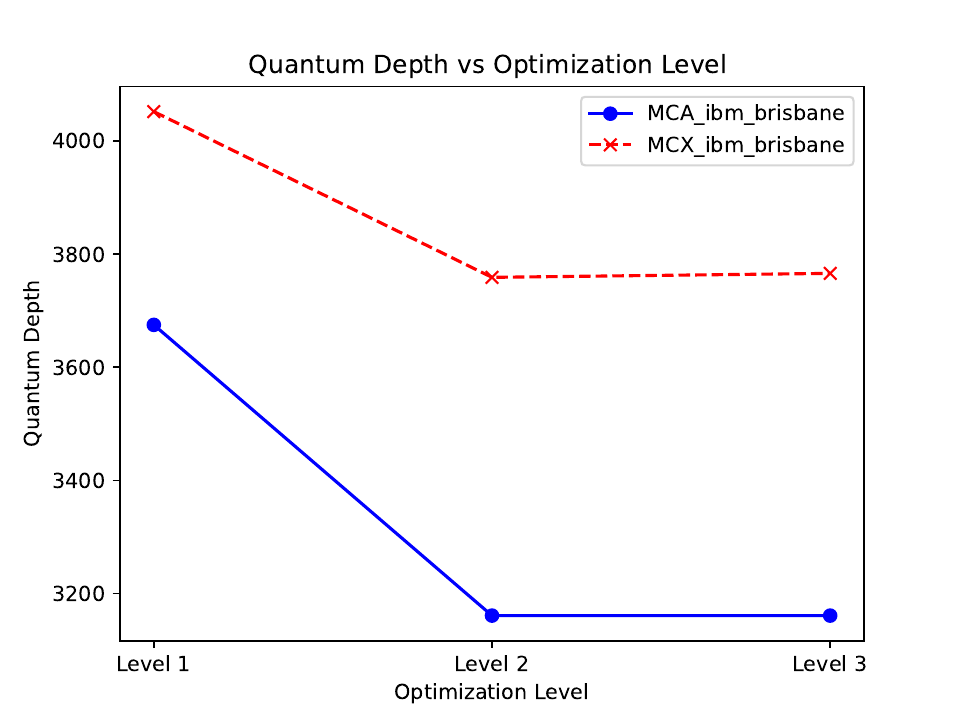}}      
{\includegraphics[width=.49\linewidth]{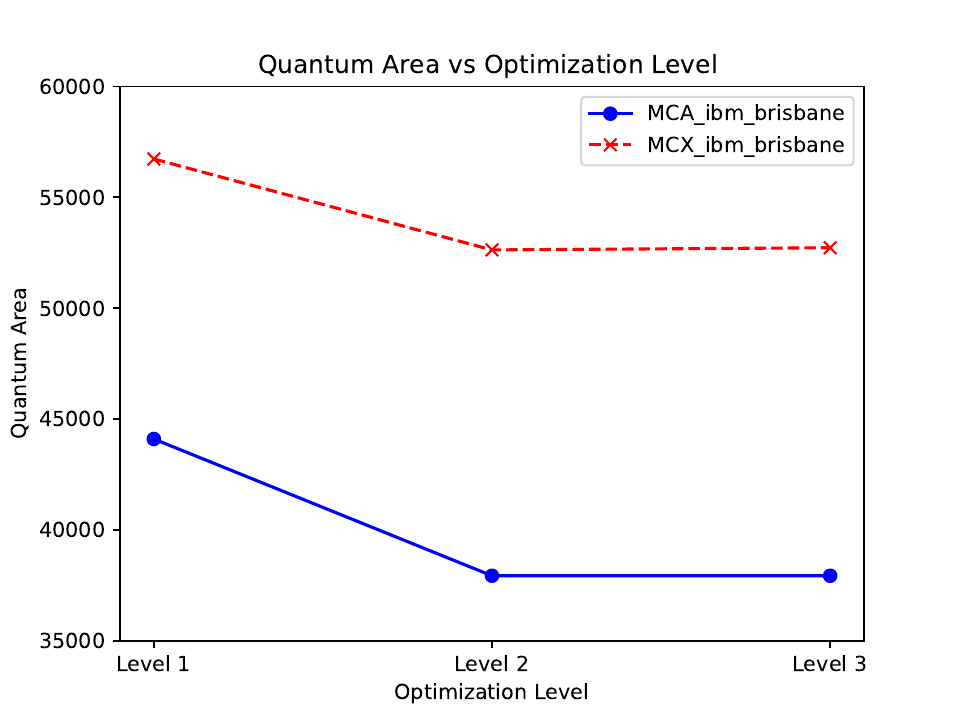}} 
\caption{Transpiled quantum circuit depth (left) and quantum cicuit area (right) for the three-eloop topology with nine edges (Fig.~\ref{fig:3e9-4e12}A) implementing the MCX and MCA quantum algorithms for different optimisation levels. }
\label{fig:depth3e9}
\end{figure}
\begin{figure}[t]
{\includegraphics[width=.49\linewidth]{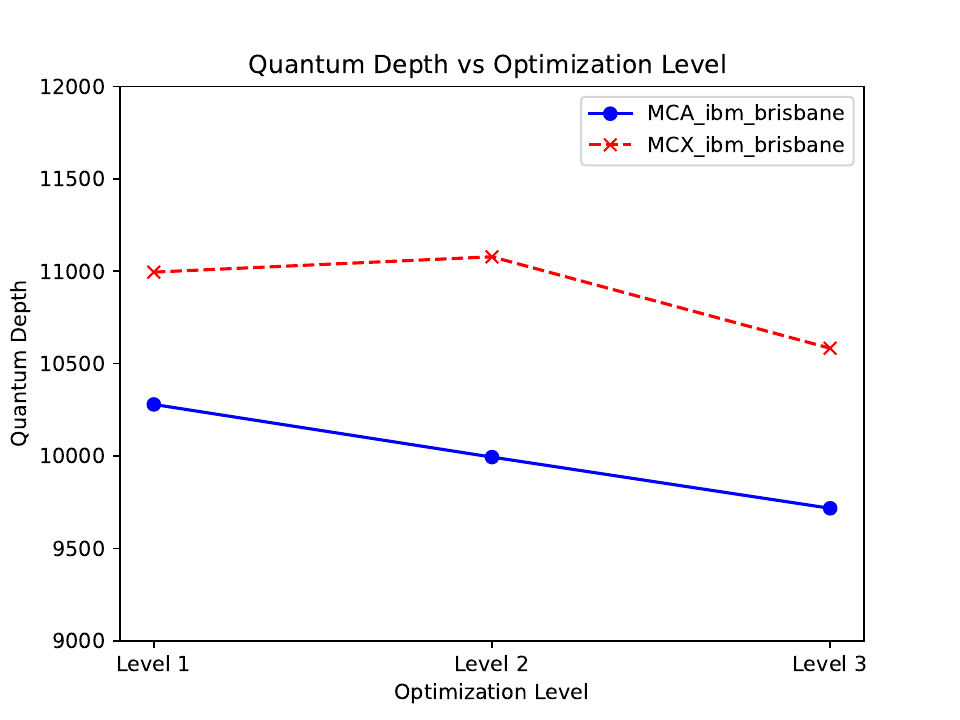}}       {\includegraphics[width=.49\linewidth]{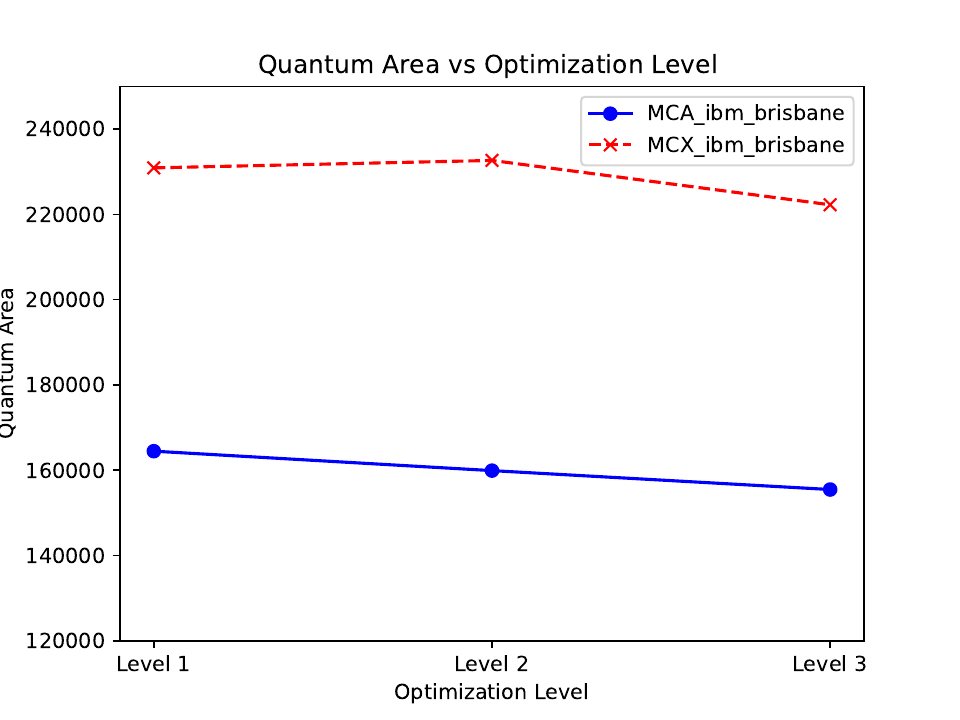}} 
\caption{Transpiled quantum circuit depth (left) and quantum cicuit area (right) for the three-eloop topology with twelve edges (Fig.\ref{fig:Sunrise_example}A) implementing the MCX and MCA quantum algorithms for different optimisation levels. }
\label{fig:depth3e12}
\end{figure}

Beyond the required number of qubits and the theoretical quantum circuit depth, we are interested in analysing the effect of transpilation, the process of compiling a given quantum circuit to match the specific topology and native gate set of a particular quantum device hardware, as well as optimising it to run on Noisy Intermediate Scale Quantum (NISQ) era computers. In order to analyse the impact of transpilation, it is important to take into account that this process may involve more qubits than those needed in the quantum circuit design. Therefore, in addition to the transpiled quantum circuit depth, we incorporate the quantum circuit area in the study. This metric considers the transpiled quantum circuit depth and the number of qubits required in the transpilation, which provides a better perspective on the computational complexity of the quantum circuit.

\begin{figure}[t]
{\includegraphics[width=.49\linewidth]{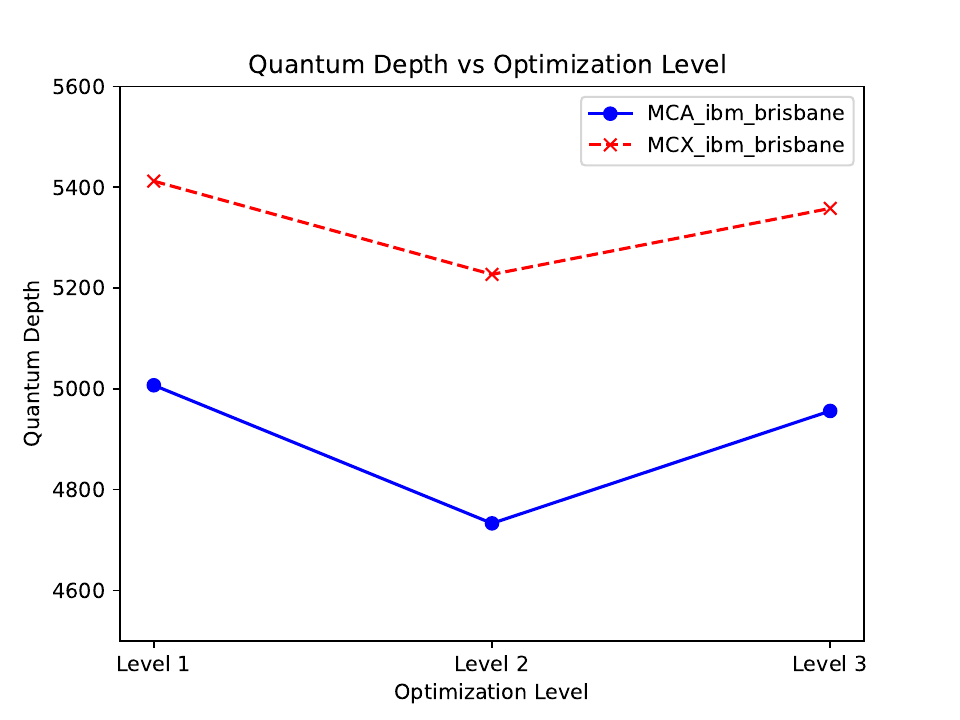}}      {\includegraphics[width=.49\linewidth]{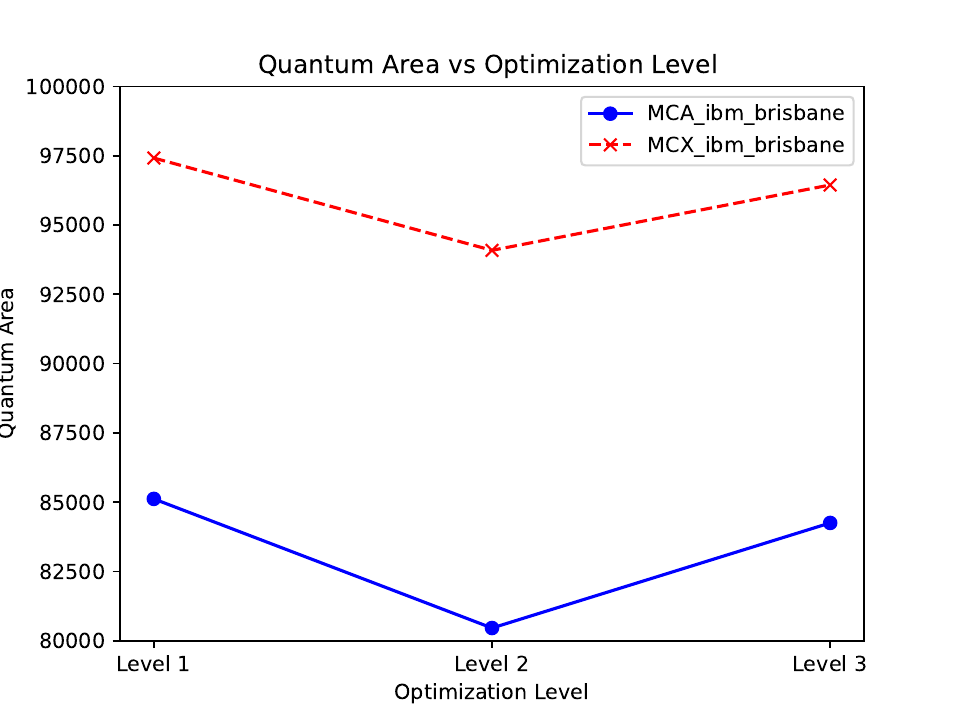}} 
\caption{Transpiled quantum circuit depth (left) and quantum circuit area (right) for the four-eloop topology with twelve edges (Fig.~\ref{fig:3e9-4e12}B) for the MCX and MCA quantum algorithms for different optimisation levels.}
\label{fig:depth4e12}
\end{figure}
\begin{figure}[t]
{\includegraphics[width=.49\linewidth]{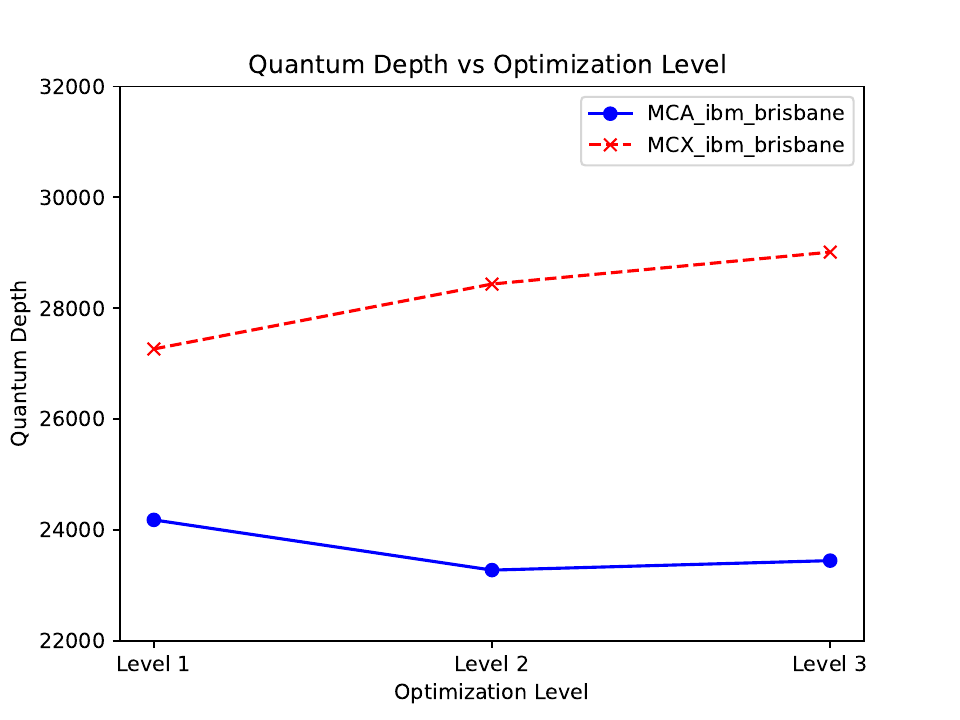}} 
{\includegraphics[width=.49\linewidth]{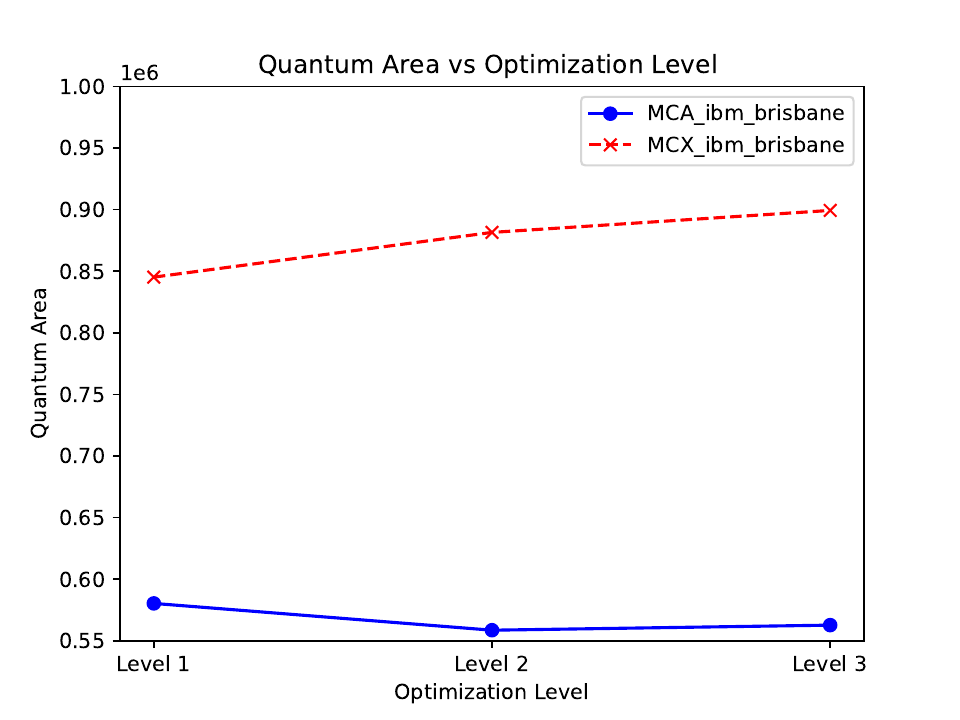}} 
\caption{Transpiled quantum circuit depth (left) and quantum circuit area (right) for the four-eloop topology with sixteen edges (Fig.~\ref{fig:3e9-4e12}C) for the MCX and MCA quantum algorithms for different optimisation levels. }
\label{depth4e16}
\end{figure}

The transpilation process is executed for relevant metrics in the available quantum backends, \texttt{ibm\_brisbane}, both simulated and real using the Qiskit framework to analize and compare the transpilation behaviour. The Qiskit framework considers different optimization levels~\cite{ibmquantumoptimization}, from zero to three, aiming to minimize the depth of the quantum circuit by reducing the noise and errors introduced by the hardware used, mainly by adapting to the topology of the quantum circuit. In the case of level zero, the Qiskit algorithm only tries to use the same number of virtual qubits as physical qubits and uses swap gates to correct the physical qubit connections caused by the hardware topology. The use of swap gates increases the quantum depth of the circuit enormously, even taking into account the transpilation of the gates used in the circuit. Therefore, we do not consider the level zero optimisation in the following analysis. Regarding the remaining levels, they apply the same steps as in level 0 and include a new method called \texttt{VF2LayoutPostLayout}~\cite{qiskit_vf2postlayout}. This method transforms the quantum circuit and the hardware into graphs, with the aim of identifying isomorphisms between them. The \texttt{VF2LayoutPostLayout} method creates a dependency on the type of hardware used, and therefore the same quantum algorithm can have a worse or better use of the physical qubits and of the quantum depth, depending on the topology of the hardware used.

We move forward to the transpilation analysis of the multiloop topologies in Fig.~\ref{fig:Sunrise_example}A and Fig.~\ref{fig:3e9-4e12}. Executing the transpilation process for the quantum circuits corresponding to the MCA and MCX algorithms applied to the three-eloop topology in Fig.~\ref{fig:3e9-4e12}A provides the transpiled quantum circuit depth and the quantum circuit area, as shown in Fig.~\ref{fig:depth3e9}. Despite the fact that no significant difference was observed in terms of the theoretical quantum circuit depth (Table~\ref{tab:comp_diag}), the MCA quantum algorithm exhibits a substantially better performance than the MCX quantum algorithm in terms of the transpiled quantum circuit depth and quantum circuit area, as shown in Fig.~\ref{fig:depth3e9}. The quantum circuit corresponding to the MCX quantum algoritm for three-eloop topology with nine edges is directly taken from Ref.~\cite{Ramirez-Uribe:2024wua}. In the case of the three-eloop topology in Fig.~\ref{fig:Sunrise_example}A, as the quantum circuit from the MCX algorithm was not presented in Ref.~\cite{Ramirez-Uribe:2024wua}, we  reconstruct it based on their methodology. The transpilation of the MCA and MCX algorithms results in the transpiled quantum circuit depth and quantum circuit area illustrated in Fig.~\ref{fig:depth3e12}. The MCA quantum algorithm also exhibits a better performance than the MCX quantum algorithm in terms of the transpiled quantum circuit depth and the quantum circuit area.

Continuing with the four-eloop topology in Fig.~\ref{fig:3e9-4e12}B, the quantum circuit from the MCX algorithm is direcly taken from Ref.~\cite{Ramirez-Uribe:2024wua} whereas the quantum circuit corresponding to the four-eloop topology in Fig.~\ref{fig:3e9-4e12}C is reconstructed. The transpiled quantum circuit depth and the quantum circuit area of the four-eloop topology with twelve edges (Fig.~\ref{fig:3e9-4e12}B) is depicted in Fig.~\ref{fig:depth4e12}, showing a favourable result for the MCA quantum algorithm. In the case of the four-eloop topology with sixteen edges (Fig.~\ref{fig:3e9-4e12}C), the transpiled quantum circuit depth in Fig.~\ref{depth4e16}~(left) is smaller with the MCA quantum algorithm. Regarding the quantum circuit area in Fig.~\ref{depth4e16}~(right), we observe that the MCA quantum algorithm outperforms MCX, which is expected due to the reduction of qubits and theoretical quantum depth.

From the analysis of the transpilation behaviour between the MCA and MCX algorithms, we have shown a significantly better performance of the MCA quantum algorithm in terms of the transpiled quantum circuit depth and the quantum circuit area. The transpiled quantum circuit depth of the MCA quantum algorithm  shows a reduction from $\mathcal{O}(7\%)$ to $\mathcal{O}(19\%)$ compared to the MCX quantum algorithm, and an improvement from $\mathcal{O}(13\%)$ to $\mathcal{O}(37\%)$ in terms of the quantum circuit area.

\section{Aplication of the MCA quantum algorithm to multiloop topologies with four and five eloops}
\label{sec:Application}

\begin{figure}[t]
    \centering  \includegraphics[width=.9\textwidth]{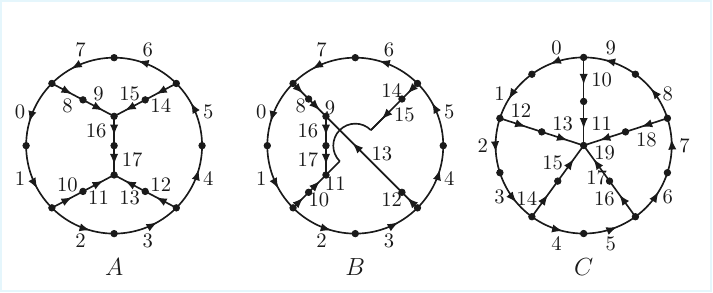} 
    \caption{From left to right: $t-$ and $u-$ channels at four eloops, and five eloops with a five-point contact interaction. Each diagram considers two edges per set.}
    \label{Topologies_figure}
\end{figure}

In this section, we implement the methodology presented in Sec.~\ref{sec:OptandAut} to the eloop topologies shown in Fig.~\ref{Topologies_figure}. These are four-eloop and five-eloop topologies whose complexity demands quantum resources that exceed the typical limits of current quantum simulators for the MCX algorithm. Therefore, we restrict the comparison to the quantum resources. The requirements for the implementation of the MCA algorithm are presented in Table~\ref{tab:Topologies_table}, including the required number of qubits and the theoretical quantum depth. For reference, Table~\ref{tab:Topologies_table} also shows the number of ancillary qubit and the total number of qubits required for the MCX quantum algorithm. 

\begin{table}[t]
\begin{tabular}{lcccccc} \hline
\textbf{Fig.} & \textbf{eloops (edges)} & \textbf{$|e\rangle$} & $|a \rangle$ & \textbf{Total Qubits} & \textbf{Quantum Depth} &\textbf{Total states}  \\
        \hline  
\ref{Topologies_figure}A & four$^{(t,s)}$ (18) & $18+1$ & $6\mid 13$ & $26\mid 33$&  $39$ & $262144$ \\ 
\ref{Topologies_figure}B & four$^{(u)}(18)$ & $18+1$ & $7\mid 15$ & $27\mid 35$ &  $49$ & $262144$ \\ 
\ref{Topologies_figure}C & five$^{(c)}$(20) & $20+1$ & $9\mid 21$ & $31\mid 43$ & $57$ & $1048576$\\ 
\hline
\end{tabular}
\caption{Quantum circuit requirements to analyse the multiloop topologies in Fig.~\ref{Topologies_figure}. The total number of qubits includes the oracle’s marker. An ancillary qubit in the $\ket{e}$ register is necessary for these multiloop topologies to achieve the optimal amplitude amplification. The first numbers in the forth and fifth columns correspond to MCA, whereas the second numbers are for MCX. The total number of qubits for MCX exceeds the typical capacity of quantum simulators, while MCA circumvent that limit.} 
\label{tab:Topologies_table}
\end{table}

\begin{figure}
    \centering   
    {\includegraphics[width=0.48\textwidth]{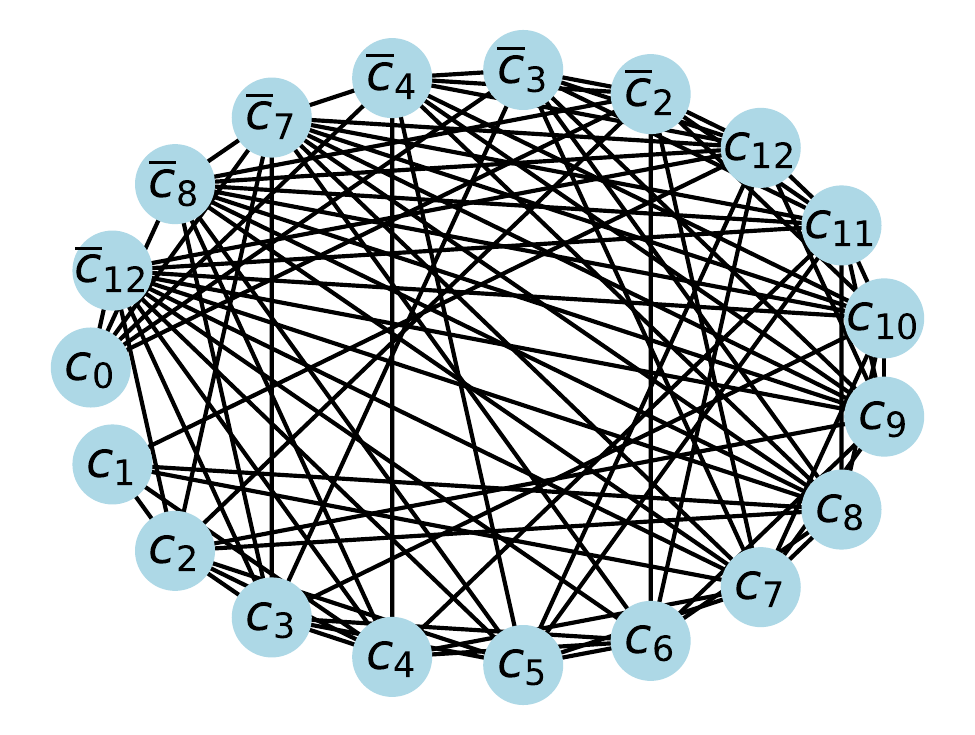}} 
    {\includegraphics[width=0.48\textwidth]{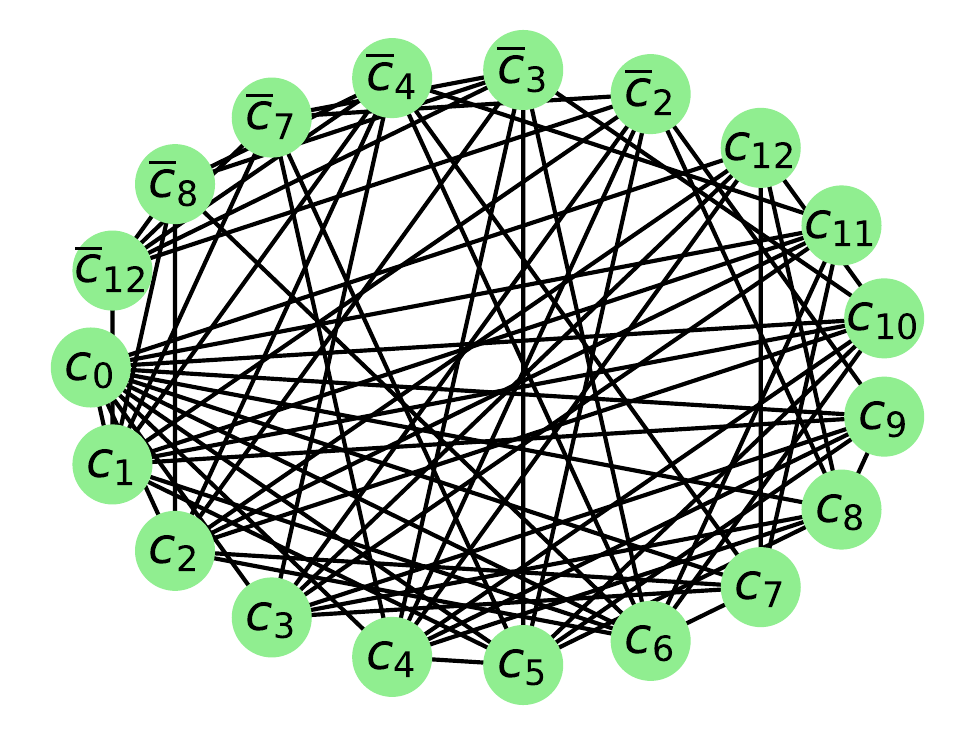}} 
\caption{Graphs generated by \texttt{MutualAuxMatrix} (left) and \texttt{MutualClauses\-Matrix} (right) by using the clauses corresponding to Fig.~\ref{Topologies_figure}A. They represent the adjacency matrix of mutually exclusive clauses, and mutually compatible clauses, respectively.}
\label{fig:Ghrap4e16}
\end{figure}

\subsection{The $t$ channel at four eloops}
\label{ssec:4eloopsst}

The $t-$channel with two edges per set is depicted in Fig.~\ref{Topologies_figure}A. With regard to the implementation of the algorithm, we tag the state of the edge $e_0$ obtaining the nineteen eloop clauses shown in Eq.~(\ref{4t18cw}). Applying the algorithms \texttt{MutualAuxMatrix} and \texttt{GraphConditionCombination} in a sequential order to \Eq{4t18cw} we obtain 
\begin{equation}
\begin{aligned}
    \texttt{MAUXc}^{(4,18)} = \{&\{\bar{c}_7, c_{8}, \bar{c}_8, c_{10}, c_{11}\}, \{ c_{0},\bar{c}_2, \bar{c}_3, \bar{c}_4\}, \{c_{6}, c_{9}, c_{12}, \bar{c}_{12}\}, \\
    &\{c_{1}, c_{2}, c_{4}\}, \{c_{3}, c_{5}\}, \{c_{7}\}\},
\end{aligned}
\end{equation}
implying the need of only six ancillary qubits in the $\ket{a}$ register. This result shows a significant reduction compared to the thirteen qubits needed in the MCX algorithm. Following with the application of the algorithms \texttt{MutualClausesMatrix}, \texttt{GraphCondition} \texttt{-Combination} and \texttt{Optuna} in a sequential order, we obtain that the optimised eloop clauses grouping is given by
\begin{equation}
\begin{aligned}
    \texttt{OMUTc}^{(4,18)} = \{ & \{c_{0}\}, \{c_{12}, c_{10}\}, \{c_{2},c_{7}, c_{6}, c_{11}, \bar{c}_{4}\}, \{\bar{c}_{8}, c_{1}, \bar{c}_{3}, \bar{c}_{12}\},\\ &\{\bar{c}_{7}\}, \{c_{4}, c_{9}, c_{5}, c_{8}, \bar{c}_{2}\}, \{c_{3}\}\}~,
\end{aligned}
\end{equation}
with theoretical quantum circuit depth of thirty-nine.

\subsection{The $u-$channel at four eloops}
\label{ssec:4eloopsu}
The $u-$channel shown in Fig.~\ref{Topologies_figure}B considers two edges per set. The total number of eloop clauses after tagging the edge $e_0$ is twenty-one, shown in Eq.~(\ref{4u18cs}). Note that the complexity of the non-planar topology gives rise to four additional eloop clauses. The implementation of the MCA quantum algorithm gives that the required ancillary qubits to store the eloop clauses are given by
\begin{equation}
\begin{aligned}
    \texttt{MAUXc}^{(4,18)} = \{&\{ c_{3}, \bar{c}_3, c_{10}, \bar{c}_{10}, c_{13}, \bar{c}_{13}\}, \{c_{2}, \bar{c}_4, \bar{c}_6, c_{11}\}, \{ \bar{c}_2, c_{4}, \bar{c}_7, c_{9}\}, \\
    &\{c_{1}, c_{6}, c_{12}\}, \{c_{0}, c_{14}\}, \{c_{5},c_{7}\}, \{c_{8}\}\}~,
\end{aligned}
\end{equation}
and the sets optimising the eloop clauses clustering in the quantum circuit are
\begin{equation}
\begin{aligned}
    \texttt{OMUTc}^{(4,18)} = \{&\{c_{4}, c_{10}, c_{12}, c_{2}\}, \{c_{0}\}, \{c_{7}, \bar{c}_{2}, c_{11}, \bar{c}_{13}, c_{8}\}, \{\bar{c}_{3}\},\\
    &\{c_{5}, \bar{c}_{4}, c_{6}, c_{13}, c_{9}\}, \{c_{3}\}, \{c_{1}, \bar{c}_{7}, c_{14}, \bar{c}_{6}, \bar{c}_{10}\}\}~.
\end{aligned}
\end{equation}
In this case, the implementation of the MCA quantum algorithm needs seven ancillary qubits in the $\ket{a}$ register, representing a reduction of $\mathcal{O}(53\%)$ compared to the fifteen ancillary qubits required in the MCX algorithm. The theoretical quantum circuit depth is forty-nine.

\subsection{Five eloops with contact interaction}
\label{ssec:5eloopscs}

The application of the MCA quantum algorithm to the five-eloop topology considering two edges per set (Fig.~\ref{Topologies_figure}C) provides the sets of eloop clauses associated to the ancillary qubits and the sets optimising the eloop clauses grouping in the quantum circuit implementation, explicitly 
\begin{equation}
\begin{aligned}
    \texttt{MAUXc}^{(5,20)} = ~ \{&\{ \bar{c}_{6}, \bar{c}_{8},c_{11}, c_{14}, c_{17}\}, \{\bar{c}_{1}, \bar{c}_{2}, c_{10},c_{16}\}, \{\bar{c}_3, \bar{c}_{4}, c_{12}, c_{18}\},
    \{\bar{c}_7, c_{8}, c_{15}, c_{19}\},\\
    &\{c_1, c_2, \bar{c}_{11}, c_{13}\}, \{ c_{3}, c_6, c_{9}, \bar{c}_{12} \},
     \{c_{4}, c_{5}, c_{7}, \bar{c}_{16}\}, \{c_{0}\}, \{c_{20}\} \},
\end{aligned}\label{eq:MAUXc5c2}
\end{equation}
and 
\begin{equation}
\begin{aligned}
    \texttt{OMUTc}^{(5,20)} = \{&\{c_{17}, c_{3}, \bar{c}_{1}, c_{12}, c_{8},c_{7}, c_{13}\},  \{\bar{c}_{6}, \bar{c}_{11}, \bar{c}_{16}, \bar{c}_{12}, \bar{c}_7, \bar{c}_{2}\},\\ &\{c_{9}, c_{14}, c_{19}, c_{0}, c_{5}, \bar{c}_{3},c_{10}\},  \{\bar{c}_{8}\}, \{\bar{c}_{4}, c_{6}, c_{15}, c_{1}, c_{11}\},\\
    &\{ c_{16}, c_{2}, c_{4}\},\{c_{20}\},\{c_{18}\}\}.
\end{aligned}
\end{equation}
Based on \Eq{eq:MAUXc5c2}, we assign nine ancillary qubits to store the eloop clauses in the MCA quantum algorithm, whereas the MCX quantum algorithm needs twenty-one. This comparison represents a reduction of $\mathcal{O}(57\%)$. Additionally, the theoretical quantum circuit depth is fifty-seven.

\section{Conclusions}
\label{sec:Conclusion}

We have presented an automated query quantum algorithm, the Minimun Clique-optimised quantum Algorithm (MCA), for the identification of DAG configurations of multiloop graphs, which is equivalent to determining the causal configurations of multiloop Feynman diagrams in particle physics. This approach introduces an automated method for oracle design by leveraging the MCP problem from graph theory in combination with hyperparameter optimisation techniques. The MCA quantum algorithm optimises both the required number of ancillary qubits and the theoretical quantum circuit depth. It has been implemented on three-, four-, and five-eloop topologies, successfully identifying the corresponding DAG or causal configurations in each case. The results demonstrate a significant reduction in the number of ancillary qubits required as the complexity of the eloop topology increases in the MCA quantum algorithm compared to the MCX quantum algorithm. Furthermore, by applying the same principles to minimise ancillary qubit usage and utilizing the Optuna framework for hyperparameter optimisation, we also achieved a considerable reduction in the theoretical quantum circuit depth.

We also evaluated the performance of quantum circuits generated by the MCA algorithm after transpilation for execution on real quantum hardware. Notably, the MCA algorithm demonstrated substantial improvements in quantum metrics, specifically in terms of reduced transpiled quantum circuit depth and smaller quantum circuit area. These results highlight the enhanced performance and resource efficiency of the MCA approach over the MCX quantum algorithm. Moreover, they also extend far beyond particle physics, establishing a use case on quantum optimization for graph theory problems and any other clause-satisfiability applications.

\section*{Acknowledgements}

This work is supported by the Mexican Government - Secretaría de Ciencia, Humanidades, Tecnología e Innovación (SECIHTI), Grants No. CBF2023-2024-268 and No. CBF2023-2024-3544 and {\it Sistema Nacional de Investigadoras e Investigadores}; the Spanish Government and ERDF/EU - Agencia Estatal de Investigaci\'on (MCIU/AEI/10.13039/501100011033), Grants No. PID2023-146220NB-I00, No. PID2020-114473GB-I00, No. EUR2025-164820, and No. CEX2023-001292-S; and Generalitat Valenciana, Grant No. ASFAE/2022/009 (Planes Complementarios de I+D+i, NextGenerationEU). This work is also supported by the Ministry of Economic Affairs and Digital Transformation of the Spanish Government and NextGenerationEU through the Quantum Spain project, and by CSIC Interdisciplinary Thematic Platform (PTI+) on Quantum Technologies (PTI-QTEP+). In addition, the work that led to these results received the support from the ``la Caixa” Foundation (ID 100010434) fellowship code is LCF/BQ/DFI25/130000.

\bibliographystyle{JHEP}
\bibliography{main}

\providecommand{\href}[2]{#2}\begingroup\raggedright\begin{thebibliography}{10}

\bibitem{Feynman:1981tf}
R.~P. Feynman, \emph{{Simulating physics with computers}},
  \href{http://dx.doi.org/10.1007/BF02650179}{\emph{Int. J. Theor. Phys.} {\bf
  21} (1982) 467--488}.

\bibitem{Humble:2022klb}
T.~S. Humble, G.~N. Perdue and M.~J. Savage, \emph{{Snowmass Computational
  Frontier: Topical Group Report on Quantum Computing}},
  \href{http://arxiv.org/abs/2209.06786}{{\tt 2209.06786}}.

\bibitem{Rodrigo:2024say}
G.~Rodrigo, \emph{{Quantum Algorithms in Particle Physics}},
  \href{http://dx.doi.org/10.5506/APhysPolBSupp.17.2-A14}{\emph{Acta Phys.
  Polon. Supp.} {\bf 17} (2024) 2--A14},
  [\href{http://arxiv.org/abs/2401.16208}{{\tt 2401.16208}}].

\bibitem{Jordan:2011ne}
S.~P. Jordan, K.~S.~M. Lee and J.~Preskill, \emph{{Quantum Algorithms for
  Quantum Field Theories}},
  \href{http://dx.doi.org/10.1126/science.1217069}{\emph{Science} {\bf 336}
  (2012) 1130--1133}, [\href{http://arxiv.org/abs/1111.3633}{{\tt 1111.3633}}].

\bibitem{Banuls:2019bmf}
M.~C. Ba\~nuls et~al., \emph{{Simulating Lattice Gauge Theories within Quantum
  Technologies}},
  \href{http://dx.doi.org/10.1140/epjd/e2020-100571-8}{\emph{Eur. Phys. J. D}
  {\bf 74} (2020) 165}, [\href{http://arxiv.org/abs/1911.00003}{{\tt
  1911.00003}}].

\bibitem{Zohar:2015hwa}
E.~Zohar, J.~I. Cirac and B.~Reznik, \emph{{Quantum Simulations of Lattice
  Gauge Theories using Ultracold Atoms in Optical Lattices}},
  \href{http://dx.doi.org/10.1088/0034-4885/79/1/014401}{\emph{Rept. Prog.
  Phys.} {\bf 79} (2016) 014401}, [\href{http://arxiv.org/abs/1503.02312}{{\tt
  1503.02312}}].

\bibitem{Byrnes:2005qx}
T.~Byrnes and Y.~Yamamoto, \emph{{Simulating lattice gauge theories on a
  quantum computer}},
  \href{http://dx.doi.org/10.1103/PhysRevA.73.022328}{\emph{Phys. Rev. A} {\bf
  73} (2006) 022328}, [\href{http://arxiv.org/abs/quant-ph/0510027}{{\tt
  quant-ph/0510027}}].

\bibitem{Magano:2021jzd}
D.~Magano et~al., \emph{{Quantum speedup for track reconstruction in particle
  accelerators}},
  \href{http://dx.doi.org/10.1103/PhysRevD.105.076012}{\emph{Phys. Rev. D} {\bf
  105} (2022) 076012}, [\href{http://arxiv.org/abs/2104.11583}{{\tt
  2104.11583}}].

\bibitem{Duckett:2022ccc}
P.~Duckett, G.~Facini, M.~Jastrzebski, S.~Malik, T.~Scanlon and S.~Rettie,
  \emph{{Reconstructing charged particle track segments with a quantum-enhanced
  support vector machine}},
  \href{http://dx.doi.org/10.1103/PhysRevD.109.052002}{\emph{Phys. Rev. D} {\bf
  109} (2024) 052002}, [\href{http://arxiv.org/abs/2212.07279}{{\tt
  2212.07279}}].

\bibitem{Schwagerl:2023elf}
T.~Schw\"agerl, C.~Issever, K.~Jansen, T.~J. Khoo, S.~K\"uhn, C.~T\"uys\"uz
  et~al., \emph{{Particle track reconstruction with noisy intermediate-scale
  quantum computers}},  \href{http://arxiv.org/abs/2303.13249}{{\tt
  2303.13249}}.

\bibitem{Wei:2019rqy}
A.~Y. Wei, P.~Naik, A.~W. Harrow and J.~Thaler, \emph{{Quantum Algorithms for
  Jet Clustering}},
  \href{http://dx.doi.org/10.1103/PhysRevD.101.094015}{\emph{Phys. Rev. D} {\bf
  101} (2020) 094015}, [\href{http://arxiv.org/abs/1908.08949}{{\tt
  1908.08949}}].

\bibitem{Pires:2020urc}
D.~Pires, Y.~Omar and J.~a. Seixas, \emph{{Adiabatic quantum algorithm for
  multijet clustering in high energy physics}},
  \href{http://dx.doi.org/10.1016/j.physletb.2023.138000}{\emph{Phys. Lett. B}
  {\bf 843} (2023) 138000}, [\href{http://arxiv.org/abs/2012.14514}{{\tt
  2012.14514}}].

\bibitem{deLejarza:2022bwc}
J.~J.~M. de~Lejarza, L.~Cieri and G.~Rodrigo, \emph{{Quantum clustering and jet
  reconstruction at the LHC}},
  \href{http://dx.doi.org/10.1103/PhysRevD.106.036021}{\emph{Phys. Rev. D} {\bf
  106} (2022) 036021}, [\href{http://arxiv.org/abs/2204.06496}{{\tt
  2204.06496}}].

\bibitem{deLejarza:2022vhe}
J.~J.~M. de~Lejarza, L.~Cieri and G.~Rodrigo, \emph{{Quantum jet clustering
  with LHC simulated data}},
  \href{http://dx.doi.org/10.22323/1.414.0241}{\emph{PoS} {\bf ICHEP2022}
  (2022) 241}, [\href{http://arxiv.org/abs/2209.08914}{{\tt 2209.08914}}].

\bibitem{Barata:2021yri}
J.~a. Barata and C.~A. Salgado, \emph{{A quantum strategy to compute the jet
  quenching parameter $\hat{q}$}},
  \href{http://dx.doi.org/10.1140/epjc/s10052-021-09674-9}{\emph{Eur. Phys. J.
  C} {\bf 81} (2021) 862}, [\href{http://arxiv.org/abs/2104.04661}{{\tt
  2104.04661}}].

\bibitem{Barata:2022wim}
J.~a. Barata, X.~Du, M.~Li, W.~Qian and C.~A. Salgado, \emph{{Medium induced
  jet broadening in a quantum computer}},
  \href{http://dx.doi.org/10.1103/PhysRevD.106.074013}{\emph{Phys. Rev. D} {\bf
  106} (2022) 074013}, [\href{http://arxiv.org/abs/2208.06750}{{\tt
  2208.06750}}].

\bibitem{Barata:2023clv}
J.~a. Barata, X.~Du, M.~Li, W.~Qian and C.~A. Salgado, \emph{{Quantum
  simulation of in-medium QCD jets: Momentum broadening, gluon production, and
  entropy growth}},
  \href{http://dx.doi.org/10.1103/PhysRevD.108.056023}{\emph{Phys. Rev. D} {\bf
  108} (2023) 056023}, [\href{http://arxiv.org/abs/2307.01792}{{\tt
  2307.01792}}].

\bibitem{Bauer:2019qxa}
C.~W. Bauer, W.~A. de~Jong, B.~Nachman and D.~Provasoli, \emph{{Quantum
  Algorithm for High Energy Physics Simulations}},
  \href{http://dx.doi.org/10.1103/PhysRevLett.126.062001}{\emph{Phys. Rev.
  Lett.} {\bf 126} (2021) 062001}, [\href{http://arxiv.org/abs/1904.03196}{{\tt
  1904.03196}}].

\bibitem{Bauer:2021gup}
C.~W. Bauer, M.~Freytsis and B.~Nachman, \emph{{Simulating Collider Physics on
  Quantum Computers Using Effective Field Theories}},
  \href{http://dx.doi.org/10.1103/PhysRevLett.127.212001}{\emph{Phys. Rev.
  Lett.} {\bf 127} (2021) 212001}, [\href{http://arxiv.org/abs/2102.05044}{{\tt
  2102.05044}}].

\bibitem{Bepari:2020xqi}
K.~Bepari, S.~Malik, M.~Spannowsky and S.~Williams, \emph{{Towards a quantum
  computing algorithm for helicity amplitudes and parton showers}},
  \href{http://dx.doi.org/10.1103/PhysRevD.103.076020}{\emph{Phys. Rev. D} {\bf
  103} (2021) 076020}, [\href{http://arxiv.org/abs/2010.00046}{{\tt
  2010.00046}}].

\bibitem{Williams:2021lvr}
K.~Bepari, S.~Malik, M.~Spannowsky and S.~Williams, \emph{{Quantum walk
  approach to simulating parton showers}},
  \href{http://dx.doi.org/10.1103/PhysRevD.106.056002}{\emph{Phys. Rev. D} {\bf
  106} (2022) 056002}, [\href{http://arxiv.org/abs/2109.13975}{{\tt
  2109.13975}}].

\bibitem{Perez-Salinas:2020nem}
A.~P\'erez-Salinas, J.~Cruz-Martinez, A.~A. Alhajri and S.~Carrazza,
  \emph{{Determining the proton content with a quantum computer}},
  \href{http://dx.doi.org/10.1103/PhysRevD.103.034027}{\emph{Phys. Rev. D} {\bf
  103} (2021) 034027}, [\href{http://arxiv.org/abs/2011.13934}{{\tt
  2011.13934}}].

\bibitem{Cruz-Martinez:2023vgs}
J.~M. Cruz-Martinez, M.~Robbiati and S.~Carrazza, \emph{{Multi-variable
  integration with a variational quantum circuit}},
  \href{http://arxiv.org/abs/2308.05657}{{\tt 2308.05657}}.

\bibitem{Chawdhry:2023jks}
H.~A. Chawdhry and M.~Pellen, \emph{{Quantum simulation of colour in
  perturbative quantum chromodynamics}},
  \href{http://dx.doi.org/10.21468/SciPostPhys.15.5.205}{\emph{SciPost Phys.}
  {\bf 15} (2023) 205}, [\href{http://arxiv.org/abs/2303.04818}{{\tt
  2303.04818}}].

\bibitem{deJong:2020tvx}
W.~A. De~Jong, M.~Metcalf, J.~Mulligan, M.~P\l{}osko\'n, F.~Ringer and X.~Yao,
  \emph{{Quantum simulation of open quantum systems in heavy-ion collisions}},
  \href{http://dx.doi.org/10.1103/PhysRevD.104.L051501}{\emph{Phys. Rev. D}
  {\bf 104} (2021) 051501}, [\href{http://arxiv.org/abs/2010.03571}{{\tt
  2010.03571}}].

\bibitem{Guan:2020bdl}
W.~Guan, G.~Perdue, A.~Pesah, M.~Schuld, K.~Terashi, S.~Vallecorsa et~al.,
  \emph{{Quantum Machine Learning in High Energy Physics}},
  \href{http://dx.doi.org/10.1088/2632-2153/abc17d}{\emph{Mach. Learn. Sci.
  Tech.} {\bf 2} (2021) 011003}, [\href{http://arxiv.org/abs/2005.08582}{{\tt
  2005.08582}}].

\bibitem{Wu:2020cye}
S.~L. Wu et~al., \emph{{Application of quantum machine learning using the
  quantum variational classifier method to high energy physics analysis at the
  LHC on IBM quantum computer simulator and hardware with 10 qubits}},
  \href{http://dx.doi.org/10.1088/1361-6471/ac1391}{\emph{J. Phys. G} {\bf 48}
  (2021) 125003}, [\href{http://arxiv.org/abs/2012.11560}{{\tt 2012.11560}}].

\bibitem{Trenti:2020ceh}
T.~Felser, M.~Trenti, L.~Sestini, A.~Gianelle, D.~Zuliani, D.~Lucchesi et~al.,
  \emph{{Quantum-inspired machine learning on high-energy physics data}},
  \href{http://dx.doi.org/10.1038/s41534-021-00443-w}{\emph{npj Quantum Inf.}
  {\bf 7} (2021) 111}, [\href{http://arxiv.org/abs/2004.13747}{{\tt
  2004.13747}}].

\bibitem{Herbert:2021xgs}
S.~Herbert, \emph{{Quantum Monte Carlo Integration: The Full Advantage in
  Minimal Circuit Depth}},
  \href{http://dx.doi.org/10.22331/q-2022-09-29-823}{\emph{Quantum} {\bf 6}
  (2022) 823}, [\href{http://arxiv.org/abs/2105.09100}{{\tt 2105.09100}}].

\bibitem{Agliardi:2022ghn}
G.~Agliardi, M.~Grossi, M.~Pellen and E.~Prati, \emph{{Quantum integration of
  elementary particle processes}},
  \href{http://dx.doi.org/10.1016/j.physletb.2022.137228}{\emph{Phys. Lett. B}
  {\bf 832} (2022) 137228}, [\href{http://arxiv.org/abs/2201.01547}{{\tt
  2201.01547}}].

\bibitem{deLejarza:2023qxk}
J.~J.~M. de~Lejarza, M.~Grossi, L.~Cieri and G.~Rodrigo, \emph{{Quantum Fourier
  Iterative Amplitude Estimation}},  in \emph{{2023 International Conference on
  Quantum Computing and Engineering}}, IEEE, 5, 2023.
\newblock \href{http://arxiv.org/abs/2305.01686}{{\tt 2305.01686}}.
\newblock \href{http://dx.doi.org/10.1109/QCE57702.2023.00071}{DOI}.

\bibitem{deLejarza:2024pgk}
J.~J.~M. de~Lejarza, L.~Cieri, M.~Grossi, S.~Vallecorsa and G.~Rodrigo,
  \emph{{Loop Feynman integration on a quantum computer}},
  \href{http://arxiv.org/abs/2401.03023}{{\tt 2401.03023}}.

\bibitem{Ramirez-Uribe:2021ubp}
S.~Ram\'\i{}rez-Uribe, A.~E. Renter\'\i{}a-Olivo, G.~Rodrigo, G.~F.~R. Sborlini
  and L.~Vale~Silva, \emph{{Quantum algorithm for Feynman loop integrals}},
  \href{http://dx.doi.org/10.1007/JHEP05(2022)100}{\emph{JHEP} {\bf 05} (2022)
  100}, [\href{http://arxiv.org/abs/2105.08703}{{\tt 2105.08703}}].

\bibitem{Clemente:2022nll}
G.~Clemente, A.~Crippa, K.~Jansen, S.~Ram\'\i{}rez-Uribe, A.~E.
  Renter\'\i{}a-Olivo, G.~Rodrigo et~al., \emph{{Variational quantum
  eigensolver for causal loop Feynman diagrams and directed acyclic graphs}},
  \href{http://dx.doi.org/10.1103/PhysRevD.108.096035}{\emph{Phys. Rev. D} {\bf
  108} (2023) 096035}, [\href{http://arxiv.org/abs/2210.13240}{{\tt
  2210.13240}}].

\bibitem{Strategy:2019vxc}
R.~K. Ellis et~al., \emph{{Physics Briefing Book}: {Input for the European
  Strategy for Particle Physics Update 2020}},
  \href{http://arxiv.org/abs/1910.11775}{{\tt 1910.11775}}.

\bibitem{Gianotti:2002xx}
F.~Gianotti et~al., \emph{{Physics potential and experimental challenges of the
  LHC luminosity upgrade}},
  \href{http://dx.doi.org/10.1140/epjc/s2004-02061-6}{\emph{Eur. Phys. J. C}
  {\bf 39} (2005) 293--333}, [\href{http://arxiv.org/abs/hep-ph/0204087}{{\tt
  hep-ph/0204087}}].

\bibitem{Abada:2019lih}
{\scshape FCC} collaboration, A.~Abada et~al., \emph{{FCC Physics
  Opportunities}: {Future Circular Collider Conceptual Design Report Volume
  1}}, \href{http://dx.doi.org/10.1140/epjc/s10052-019-6904-3}{\emph{Eur. Phys.
  J. C} {\bf 79} (2019) 474}.

\bibitem{Djouadi:2007ik}
{\scshape ILC} collaboration, G.~Aarons et~al., \emph{{International Linear
  Collider Reference Design Report Volume 2: Physics at the ILC}},
  \href{http://arxiv.org/abs/0709.1893}{{\tt 0709.1893}}.

\bibitem{Roloff:2018dqu}
{\scshape CLIC, CLICdp} collaboration, \emph{{The Compact Linear e$^+$e$^-$
  Collider (CLIC): Physics Potential}},
  \href{http://arxiv.org/abs/1812.07986}{{\tt 1812.07986}}.

\bibitem{CEPCStudyGroup:2018ghi}
{\scshape CEPC Study Group} collaboration, M.~Dong et~al., \emph{{CEPC
  Conceptual Design Report: Volume 2 - Physics \& Detector}},
  \href{http://arxiv.org/abs/1811.10545}{{\tt 1811.10545}}.

\bibitem{Heinrich:2020ybq}
G.~Heinrich, \emph{{Collider Physics at the Precision Frontier}},
  \href{http://dx.doi.org/10.1016/j.physrep.2021.03.006}{\emph{Phys. Rept.}
  {\bf 922} (2021) 1--69}, [\href{http://arxiv.org/abs/2009.00516}{{\tt
  2009.00516}}].

\bibitem{Catani:2008xa}
S.~Catani, T.~Gleisberg, F.~Krauss, G.~Rodrigo and J.-C. Winter, \emph{{From
  loops to trees by-passing Feynman's theorem}},
  \href{http://dx.doi.org/10.1088/1126-6708/2008/09/065}{\emph{JHEP} {\bf 09}
  (2008) 065}, [\href{http://arxiv.org/abs/0804.3170}{{\tt 0804.3170}}].

\bibitem{Bierenbaum:2010cy}
I.~Bierenbaum, S.~Catani, P.~Draggiotis and G.~Rodrigo, \emph{{A Tree-Loop
  Duality Relation at Two Loops and Beyond}},
  \href{http://dx.doi.org/10.1007/JHEP10(2010)073}{\emph{JHEP} {\bf 10} (2010)
  073}, [\href{http://arxiv.org/abs/1007.0194}{{\tt 1007.0194}}].

\bibitem{Bierenbaum:2012th}
I.~Bierenbaum, S.~Buchta, P.~Draggiotis, I.~Malamos and G.~Rodrigo,
  \emph{{Tree-Loop Duality Relation beyond simple poles}},
  \href{http://dx.doi.org/10.1007/JHEP03(2013)025}{\emph{JHEP} {\bf 03} (2013)
  025}, [\href{http://arxiv.org/abs/1211.5048}{{\tt 1211.5048}}].

\bibitem{Tomboulis:2017rvd}
E.~T. Tomboulis, \emph{{Causality and Unitarity via the Tree-Loop Duality
  Relation}}, \href{http://dx.doi.org/10.1007/JHEP05(2017)148}{\emph{JHEP} {\bf
  05} (2017) 148}, [\href{http://arxiv.org/abs/1701.07052}{{\tt 1701.07052}}].

\bibitem{Runkel:2019yrs}
R.~Runkel, Z.~Sz\H{o}r, J.~P. Vesga and S.~Weinzierl, \emph{{Causality and
  loop-tree duality at higher loops}},
  \href{http://dx.doi.org/10.1103/PhysRevLett.122.111603}{\emph{Phys. Rev.
  Lett.} {\bf 122} (2019) 111603}, [\href{http://arxiv.org/abs/1902.02135}{{\tt
  1902.02135}}].

\bibitem{Capatti:2019ypt}
Z.~Capatti, V.~Hirschi, D.~Kermanschah and B.~Ruijl, \emph{{Loop-Tree Duality
  for Multiloop Numerical Integration}},
  \href{http://dx.doi.org/10.1103/PhysRevLett.123.151602}{\emph{Phys. Rev.
  Lett.} {\bf 123} (2019) 151602}, [\href{http://arxiv.org/abs/1906.06138}{{\tt
  1906.06138}}].

\bibitem{Buchta:2014dfa}
S.~Buchta, G.~Chachamis, P.~Draggiotis, I.~Malamos and G.~Rodrigo, \emph{{On
  the singular behaviour of scattering amplitudes in quantum field theory}},
  \href{http://dx.doi.org/10.1007/JHEP11(2014)014}{\emph{JHEP} {\bf 11} (2014)
  014}, [\href{http://arxiv.org/abs/1405.7850}{{\tt 1405.7850}}].

\bibitem{Hernandez-Pinto:2015ysa}
R.~J. Hernandez-Pinto, G.~F.~R. Sborlini and G.~Rodrigo, \emph{{Towards gauge
  theories in four dimensions}},
  \href{http://dx.doi.org/10.1007/JHEP02(2016)044}{\emph{JHEP} {\bf 02} (2016)
  044}, [\href{http://arxiv.org/abs/1506.04617}{{\tt 1506.04617}}].

\bibitem{Jurado:2017xut}
J.~L. Jurado, G.~Rodrigo and W.~J. Torres~Bobadilla, \emph{{From Jacobi
  off-shell currents to integral relations}},
  \href{http://dx.doi.org/10.1007/JHEP12(2017)122}{\emph{JHEP} {\bf 12} (2017)
  122}, [\href{http://arxiv.org/abs/1710.11010}{{\tt 1710.11010}}].

\bibitem{Driencourt-Mangin:2019yhu}
F.~Driencourt-Mangin, G.~Rodrigo, G.~F.~R. Sborlini and W.~J. Torres~Bobadilla,
  \emph{{Interplay between the loop-tree duality and helicity amplitudes}},
  \href{http://dx.doi.org/10.1103/PhysRevD.105.016012}{\emph{Phys. Rev. D} {\bf
  105} (2022) 016012}, [\href{http://arxiv.org/abs/1911.11125}{{\tt
  1911.11125}}].

\bibitem{Aguilera-Verdugo:2019kbz}
J.~J. Aguilera-Verdugo, F.~Driencourt-Mangin, J.~Plenter,
  S.~Ram\'\i{}rez-Uribe, G.~Rodrigo, G.~F.~R. Sborlini et~al.,
  \emph{{Causality, unitarity thresholds, anomalous thresholds and infrared
  singularities from the loop-tree duality at higher orders}},
  \href{http://dx.doi.org/10.1007/JHEP12(2019)163}{\emph{JHEP} {\bf 12} (2019)
  163}, [\href{http://arxiv.org/abs/1904.08389}{{\tt 1904.08389}}].

\bibitem{Buchta:2015wna}
S.~Buchta, G.~Chachamis, P.~Draggiotis and G.~Rodrigo, \emph{{Numerical
  implementation of the loop\textendash{}tree duality method}},
  \href{http://dx.doi.org/10.1140/epjc/s10052-017-4833-6}{\emph{Eur. Phys. J.
  C} {\bf 77} (2017) 274}, [\href{http://arxiv.org/abs/1510.00187}{{\tt
  1510.00187}}].

\bibitem{Sborlini:2016gbr}
G.~F.~R. Sborlini, F.~Driencourt-Mangin, R.~Hernandez-Pinto and G.~Rodrigo,
  \emph{{Four-dimensional unsubtraction from the loop-tree duality}},
  \href{http://dx.doi.org/10.1007/JHEP08(2016)160}{\emph{JHEP} {\bf 08} (2016)
  160}, [\href{http://arxiv.org/abs/1604.06699}{{\tt 1604.06699}}].

\bibitem{Sborlini:2016hat}
G.~F.~R. Sborlini, F.~Driencourt-Mangin and G.~Rodrigo, \emph{{Four-dimensional
  unsubtraction with massive particles}},
  \href{http://dx.doi.org/10.1007/JHEP10(2016)162}{\emph{JHEP} {\bf 10} (2016)
  162}, [\href{http://arxiv.org/abs/1608.01584}{{\tt 1608.01584}}].

\bibitem{Driencourt-Mangin:2017gop}
F.~Driencourt-Mangin, G.~Rodrigo and G.~F.~R. Sborlini, \emph{{Universal dual
  amplitudes and asymptotic expansions for $gg\rightarrow H$ and $H\rightarrow
  \gamma \gamma $ in four dimensions}},
  \href{http://dx.doi.org/10.1140/epjc/s10052-018-5692-5}{\emph{Eur. Phys. J.
  C} {\bf 78} (2018) 231}, [\href{http://arxiv.org/abs/1702.07581}{{\tt
  1702.07581}}].

\bibitem{Driencourt-Mangin:2019aix}
F.~Driencourt-Mangin, G.~Rodrigo, G.~F.~R. Sborlini and W.~J. Torres~Bobadilla,
  \emph{{Universal four-dimensional representation of $H \to \gamma \gamma$ at
  two loops through the Loop-Tree Duality}},
  \href{http://dx.doi.org/10.1007/JHEP02(2019)143}{\emph{JHEP} {\bf 02} (2019)
  143}, [\href{http://arxiv.org/abs/1901.09853}{{\tt 1901.09853}}].

\bibitem{Capatti:2019edf}
Z.~Capatti, V.~Hirschi, D.~Kermanschah, A.~Pelloni and B.~Ruijl,
  \emph{{Numerical Loop-Tree Duality: contour deformation and subtraction}},
  \href{http://dx.doi.org/10.1007/JHEP04(2020)096}{\emph{JHEP} {\bf 04} (2020)
  096}, [\href{http://arxiv.org/abs/1912.09291}{{\tt 1912.09291}}].

\bibitem{Plenter:2019jyj}
J.~Plenter, \emph{{Asymptotic Expansions Through the Loop-Tree Duality}},
  \href{http://dx.doi.org/10.5506/APhysPolB.50.1983}{\emph{Acta Phys. Polon. B}
  {\bf 50} (2019) 1983--1992}.

\bibitem{Prisco:2020kyb}
R.~M. Prisco and F.~Tramontano, \emph{{Dual subtractions}},
  \href{http://dx.doi.org/10.1007/JHEP06(2021)089}{\emph{JHEP} {\bf 06} (2021)
  089}, [\href{http://arxiv.org/abs/2012.05012}{{\tt 2012.05012}}].

\bibitem{Plenter:2020lop}
J.~Plenter and G.~Rodrigo, \emph{{Asymptotic expansions through the loop-tree
  duality}},
  \href{http://dx.doi.org/10.1140/epjc/s10052-021-09094-9}{\emph{Eur. Phys. J.
  C} {\bf 81} (2021) 320}, [\href{http://arxiv.org/abs/2005.02119}{{\tt
  2005.02119}}].

\bibitem{Runkel:2019zbm}
R.~Runkel, Z.~Sz\H{o}r, J.~P. Vesga and S.~Weinzierl, \emph{{Integrands of loop
  amplitudes within loop-tree duality}},
  \href{http://dx.doi.org/10.1103/PhysRevD.101.116014}{\emph{Phys. Rev. D} {\bf
  101} (2020) 116014}, [\href{http://arxiv.org/abs/1906.02218}{{\tt
  1906.02218}}].

\bibitem{Verdugo:2020kzh}
J.~J. Aguilera-Verdugo, F.~Driencourt-Mangin, R.~J. Hern\'andez-Pinto,
  J.~Plenter, S.~Ramirez-Uribe, A.~E. Renteria~Olivo et~al., \emph{{Open Loop
  Amplitudes and Causality to All Orders and Powers from the Loop-Tree
  Duality}},
  \href{http://dx.doi.org/10.1103/PhysRevLett.124.211602}{\emph{Phys. Rev.
  Lett.} {\bf 124} (2020) 211602}, [\href{http://arxiv.org/abs/2001.03564}{{\tt
  2001.03564}}].

\bibitem{snowmass2020}
J.~J. Aguilera-Verdugo, R.~J. Hern\'andez-Pinto, S.~Ram\'\i{}rez-Uribe, A.~E.
  Renter\'\i{}a-Olivo, G.~Rodrigo, G.~F.~R. Sborlini et~al., \emph{{Manifestly
  Causal Scattering Amplitudes}},
  \href{http://dx.doi.org/https://www.snowmass21.org/docs/files/summaries/TF/SNOWMASS21-TF4_TF6_TorresBobadilla-093.pdf}{\emph{Snowmass
  LoI} (August 2020) }.

\bibitem{Aguilera-Verdugo:2020kzc}
J.~J. Aguilera-Verdugo, R.~J. Hernandez-Pinto, G.~Rodrigo, G.~F.~R. Sborlini
  and W.~J. Torres~Bobadilla, \emph{{Causal representation of multi-loop
  Feynman integrands within the loop-tree duality}},
  \href{http://dx.doi.org/10.1007/JHEP01(2021)069}{\emph{JHEP} {\bf 01} (2021)
  069}, [\href{http://arxiv.org/abs/2006.11217}{{\tt 2006.11217}}].

\bibitem{Aguilera-Verdugo:2020nrp}
J.~Jes\'us Aguilera-Verdugo, R.~J. Hern\'andez-Pinto, G.~Rodrigo, G.~F.~R.
  Sborlini and W.~J. Torres~Bobadilla, \emph{{Mathematical properties of nested
  residues and their application to multi-loop scattering amplitudes}},
  \href{http://dx.doi.org/10.1007/JHEP02(2021)112}{\emph{JHEP} {\bf 02} (2021)
  112}, [\href{http://arxiv.org/abs/2010.12971}{{\tt 2010.12971}}].

\bibitem{Ramirez-Uribe:2020hes}
S.~Ram\'\i{}rez-Uribe, R.~J. Hern\'andez-Pinto, G.~Rodrigo, G.~F.~R. Sborlini
  and W.~J. Torres~Bobadilla, \emph{{Universal opening of four-loop scattering
  amplitudes to trees}},
  \href{http://dx.doi.org/10.1007/JHEP04(2021)129}{\emph{JHEP} {\bf 04} (2021)
  129}, [\href{http://arxiv.org/abs/2006.13818}{{\tt 2006.13818}}].

\bibitem{Sborlini:2021owe}
G.~F.~R. Sborlini, \emph{{Geometrical approach to causality in multiloop
  amplitudes}},
  \href{http://dx.doi.org/10.1103/PhysRevD.104.036014}{\emph{Phys. Rev. D} {\bf
  104} (2021) 036014}, [\href{http://arxiv.org/abs/2102.05062}{{\tt
  2102.05062}}].

\bibitem{TorresBobadilla:2021ivx}
W.~J. Torres~Bobadilla, \emph{{Loop-tree duality from vertices and edges}},
  \href{http://dx.doi.org/10.1007/JHEP04(2021)183}{\emph{JHEP} {\bf 04} (2021)
  183}, [\href{http://arxiv.org/abs/2102.05048}{{\tt 2102.05048}}].

\bibitem{TorresBobadilla:2021dkq}
W.~J.~T. Bobadilla, \emph{{Lotty \textendash{} The loop-tree duality
  automation}},
  \href{http://dx.doi.org/10.1140/epjc/s10052-021-09235-0}{\emph{Eur. Phys. J.
  C} {\bf 81} (2021) 514}, [\href{http://arxiv.org/abs/2103.09237}{{\tt
  2103.09237}}].

\bibitem{Aguilera-Verdugo:2021nrn}
J.~de~Jes\'us Aguilera-Verdugo et~al., \emph{{A Stroll through the Loop-Tree
  Duality}}, \href{http://dx.doi.org/10.3390/sym13061029}{\emph{Symmetry} {\bf
  13} (2021) 1029}, [\href{http://arxiv.org/abs/2104.14621}{{\tt 2104.14621}}].

\bibitem{Rios-Sanchez:2024xtv}
J.~Rios-Sanchez and G.~Sborlini, \emph{{Towards multiloop local renormalization
  within Causal Loop-Tree Duality}},
  \href{http://arxiv.org/abs/2402.13995}{{\tt 2402.13995}}.

\bibitem{EPpatent}
S.~Ram\'\i{}rez-Uribe, A.~E. Renter\'\i{}a-Olivo, G.~Rodrigo and G.~F.~R.
  Sborlini, \emph{{Method for generating in a random order directed acyclic
  graph (DAG) configurations of a given graph in a quantum computing device }},
  {\emph{European patent application EP24382213} (28 February 2024) }.

\bibitem{Bollini:1972ui}
C.~G. Bollini and J.~J. Giambiagi, \emph{{Dimensional Renormalization: The
  Number of Dimensions as a Regularizing Parameter}},
  \href{http://dx.doi.org/10.1007/BF02895558}{\emph{Nuovo Cim. B} {\bf 12}
  (1972) 20--26}.

\bibitem{tHooft:1972tcz}
G.~'t~Hooft and M.~J.~G. Veltman, \emph{{Regularization and Renormalization of
  Gauge Fields}},
  \href{http://dx.doi.org/10.1016/0550-3213(72)90279-9}{\emph{Nucl. Phys. B}
  {\bf 44} (1972) 189--213}.

\bibitem{Boyer:1996zf}
M.~Boyer, G.~Brassard, P.~Hoeyer and A.~Tapp, \emph{{Tight bounds on quantum
  searching}},
  \href{http://dx.doi.org/10.1002/(SICI)1521-3978(199806)46:4/5<493::AID-PROP493>3.0.CO;2-P}{\emph{Fortsch.
  Phys.} {\bf 46} (1998) 493--506},
  [\href{http://arxiv.org/abs/quant-ph/9605034}{{\tt quant-ph/9605034}}].

\bibitem{Grover:1997fa}
L.~K. Grover, \emph{{Quantum mechanics helps in searching for a needle in a
  haystack}}, \href{http://dx.doi.org/10.1103/PhysRevLett.79.325}{\emph{Phys.
  Rev. Lett.} {\bf 79} (1997) 325--328},
  [\href{http://arxiv.org/abs/quant-ph/9706033}{{\tt quant-ph/9706033}}].

\bibitem{Nielsen:2012yss}
M.~A. Nielsen and I.~L. Chuang, \emph{{Quantum Computation and Quantum
  Information}},
  \href{http://dx.doi.org/10.1017/cbo9780511976667}{\emph{Cambridge University
  Press} (6, 2012) }.

\bibitem{Depth:5769}
\emph{https://quantumcomputing.stackexchange.com/questions/5769/how-to-calculate-circuit-depth-properly},
  2020.

\end{thebibliography}\endgroup

\appendix
\section{Clauses of multiloop topologies: four and five eloops}

\subsection{Four eloops: $s$- and $t$-channels}

The corresponding eloop clauses for the $t$-channel with two edges per set shown in Fig.~\ref{Topologies_figure}A are 
\beq
\begin{aligned}
c_0  &= s_0 \wedge s_1 \wedge s_2 \wedge s_3~, 
& c_1  &= s_0 \wedge \bar{s}_4 \wedge s_5 ~, \\
c_2  &= s_1 \wedge \bar{s}_5 \wedge s_6 \wedge s_8~, 
& c_3  &= s_2 \wedge \bar{s}_6 \wedge s_7 ~,\\
c_4  &=s_3 \wedge s_4 \wedge \bar{s}_7 \wedge \bar{s}_8 ~, 
& c_5  &=s_0 \wedge s_3 \wedge s_5 \wedge \bar{s}_7 \wedge \bar{s}_8~, \\
c_6  &= s_0 \wedge s_1 \wedge \bar{s}_4 \wedge s_6 \wedge s_8~, 
& c_7  &=  s_1 \wedge s_2 \wedge \bar{s}_5  \wedge s_7 \wedge s_8~,\\
c_8  &= s_2  \wedge s_3 \wedge s_4 \wedge \bar{s}_6 \wedge \bar{s}_8~, 
& c_9  & = s_0 \wedge s_2 \wedge s_3 \wedge s_5 \wedge \bar{s}_6 \wedge \bar{s}_8~, \\
c_{10}  & = s_0 \wedge s_1 \wedge s_3 \wedge s_6 \wedge \bar{s}_7~, 
& c_{11}  &= s_0  \wedge s_1 \wedge s_2 \wedge \bar{s}_4 \wedge s_7 \wedge s_8~, \\
c_{12}  &= s_1 \wedge s_2 \wedge s_3 \wedge s_4 \wedge \bar{s}_5 ~, 
& c_{13}  &= \bar{c}_2~,\qquad c_{14} = \bar{c}_3~, \\
c_{15}  &=\bar{c}_4~, \qquad c_{16} = \bar{c}_7~, \qquad 
& c_{17}  &= \bar{c}_8~, \qquad c_{18}  =\bar{c}_{12}~. 
\end{aligned}\\
\label{4t18cw}
\eeq
The eloop clauses for the so-called $s$-channel~\cite{Ramirez-Uribe:2020hes} correspond to a proper rotation and relabeling of the sets of edges.

\subsection{Four eloops: $u$-channel}

The corresponding eloop clauses for the $u$-channel with two edges per set shown in Fig.~\ref{Topologies_figure}B are
\beq
\begin{gathered}
\begin{aligned}
c_0  &= s_0 \wedge s_1 \wedge s_2 \wedge s_3~, 
& c_1  &= s_0 \wedge \bar{s}_4 \wedge s_5 \wedge \bar{s}_8~, \\
c_2  &= s_1 \wedge \bar{s}_5 \wedge s_6 \wedge s_8~,
& c_3  &= s_2 \wedge \bar{s}_6 \wedge s_7 \wedge \bar{s}_8~,\\
c_4  &=s_3 \wedge s_4 \wedge \bar{s}_7 \wedge s_8~, 
& c_5  &=s_0 \wedge s_1 \wedge \bar{s}_4 \wedge s_6~, \\
c_6  &=s_1 \wedge s_2 \wedge \bar{s}_5 \wedge s_7~, 
& c_7  &=  s_2 \wedge s_3  \wedge s_4 \wedge\bar{s}_6~,\\
c_8  &= s_0  \wedge s_3 \wedge s_5 \wedge \bar{s}_7~, 
& c_9  &= s_0 \wedge s_1 \wedge s_2 \wedge \bar{s}_4 \wedge s_7 \wedge \bar{s}_8~, \\
c_{10}  &=s_1 \wedge s_2 \wedge s_3 \wedge s_4 \wedge \bar{s}_5 \wedge s_8~, 
& c_{11}  &= s_0  \wedge s_2 \wedge s_3 \wedge s_5 \wedge \bar{s}_6 \wedge \bar{s}_8~, \\
c_{12}  &= s_0 \wedge s_1 \wedge s_3 \wedge s_6 \wedge \bar{s}_7 \wedge s_8~, 
& c_{13}  &= s_1  \wedge \bar{s}_3 \wedge \bar{s}_4 \wedge \bar{s}_5 \wedge s_6 \wedge s_7~,\\
c_{14}  &= s_0 \wedge \bar{s}_2 \wedge \bar{s}_4 \wedge s_5 \wedge s_6 \wedge \bar{s}_7~, 
& c_{15} &=\bar{c}_2~, \quad c_{16} = \bar{c}_3~, \quad
c_{17} = \bar{c}_4~, \\
c_{18} &=\bar{c}_6~, \quad c_{19} = \bar{c}_7~, \quad c_{20} = \bar{c}_{10}~, 
& c_{21} &= \bar{c}_{13}~.
\end{aligned}\\
\end{gathered}
\label{4u18cs}
\eeq

\subsection{Five eloops with contact interaction}

The corresponding eloop clauses for the five-eloop topology with two edges per set shown in Fig.~\ref{Topologies_figure}C are
\beq
\begin{gathered}
\begin{aligned}
c_0  &= s_0 \wedge \bar{s}_5 \wedge s_6~, & c_1  &= s_1 \wedge \bar{s}_6 \wedge s_7~,  \\
 c_2  &= s_2 \wedge \bar{s}_7 \wedge s_8~, & c_3  &=s_3 \wedge \bar{s}_8 \wedge s_9~,\\
 c_4 &= s_4 \wedge s_5 \wedge \bar{s}_9~,& c_5  &=s_0 \wedge s_1 \wedge \bar{s}_5 \wedge s_7~, \\
 c_6  &=s_1 \wedge s_2 \wedge \bar{s}_6 \wedge s_8~, & c_7  &=s_2 \wedge s_3 \wedge \bar{s}_7 \wedge s_9~, \\
c_{8}  &=s_3 \wedge s_4 \wedge s_5 \wedge \bar{s}_8~,& c_9  &= s_0 \wedge s_4 \wedge s_6 \wedge \bar{s}_9~, \\
c_{10}  &=s_0 \wedge s_1 \wedge s_2 \wedge \bar{s}_5 \wedge s_8~,& c_{11} &= s_1 \wedge s_2 \wedge s_3 \wedge \bar{s}_6 \wedge s_9~,\\ 
c_{12} &=s_2 \wedge s_3 \wedge s_4 \wedge s_5 \wedge \bar{s}_7~, 
& c_{13}  &= s_0 \wedge s_3 \wedge s_4 \wedge s_6 \wedge \bar{s}_8~, \\ 
 c_{14} &=s_0 \wedge s_1 \wedge s_4 \wedge s_7 \wedge \bar{s}_9~,& c_{15}  &= s_0 \wedge s_1 \wedge s_2 \wedge s_3 \wedge \bar{s}_5 \wedge s_9~, \\
c_{16}&=s_1 \wedge s_2 \wedge s_3 \wedge s_4 \wedge s_5 \wedge \bar{s}_6~,& c_{17}  &=s_0 \wedge s_2 \wedge s_3 \wedge s_4 \wedge s_6 \wedge \bar{s}_7~, \\
c_{18}  &= s_0 \wedge s_1 \wedge s_3 \wedge s_4 \wedge s_7 \wedge \bar{s}_8~,
& c_{19}  &=s_0 \wedge s_1 \wedge s_2 \wedge s_4 \wedge s_8 \wedge \bar{s}_9~, \\
 c_{20}  &= s_0 \wedge s_1 \wedge s_2 \wedge s_3 \wedge s_4~,
& c_{21}  &= \bar{c}_1~,\qquad~~~~  c_{22} = \bar{c}_2~,\\
c_{23} &= \bar{c}_3~, \qquad~~~~ c_{24} = \bar{c}_4~, 
& c_{25} &= \bar{c}_6~,\qquad~~~~ c_{26} = \bar{c}_7~, \\
c_{27} &= \bar{c}_{8}~,\qquad~~ c_{28} = \bar{c}_{11}~,
& c_{29} &=\bar{c}_{12}~, \qquad~~
c_{30} = \bar{c}_{16}~.
\end{aligned}\\
\end{gathered}
\label{5e20cw}
\eeq

\end{document}